\documentclass[aps,prx,reprint,superscriptaddress,floatfix]{revtex4-2}
\usepackage{amsmath,amssymb,wasysym}
\usepackage{kotex}
\usepackage{graphicx}
\usepackage{dcolumn}
\usepackage{bm}
\usepackage{braket}
\usepackage{verbatim}
\usepackage{float}
\usepackage{bbold}
\usepackage{color}
\usepackage{xcolor}
\usepackage{relsize}
\usepackage{amsthm}
\usepackage{enumerate}
\usepackage{soul,xcolor}
\usepackage[T1]{fontenc}
\usepackage{adjustbox}
\usepackage[colorlinks=true ,urlcolor=blue,urlbordercolor={0 1 1}]{hyperref}

\newcommand{\up}{\uparrow}
\newcommand{\down}{\downarrow}

\renewcommand{\[}{\left[}

\newcommand{\bk}{\bm{k}}

\newcommand{\be}{\begin{equation}}
\newcommand{\ba}{\begin{align}}
\newcommand{\ee}{\end{equation}}
\newcommand{\bea}{\begin{eqnarray}}
\newcommand{\eea}{\end{eqnarray}}
\newcommand{\beq}{\begin{equation}}
\newcommand{\eeq}{\end{equation}}
\newcommand{\beqn}{\begin{eqnarray}}
\newcommand{\eeqn}{\end{eqnarray}}

\renewcommand{\hat}[1]{{\widehat #1}}

\usepackage{bm}
\usepackage{array}
\newcolumntype{L}[1]{>{\raggedright\arraybackslash}p{#1}}
\newcolumntype{C}[1]{>{\centering\arraybackslash}p{#1}}
\newcolumntype{R}[1]{>{\raggedleft\arraybackslash}p{#1}}
\usepackage{multirow}
\allowdisplaybreaks

\newcolumntype{C}[1]{>{\centering\arraybackslash}p{#1}}

\usepackage{pifont}
\usepackage{tikz}
\usetikzlibrary{decorations.pathmorphing,decorations.markings,arrows.meta}

\usetikzlibrary{decorations.markings,arrows.meta,bending}
\usetikzlibrary{decorations.pathmorphing}   
\tikzset{
  arc arrow/.style args={to pos #1 with length #2}{
    decoration={
      markings,
      mark=at position 0 with {\pgfextra{%
        \pgfmathsetmacro{\arclen}{#2}%
        \pgfmathsetmacro{\arrowtime}{\arclen/\pgfdecoratedpathlength}%
        \xdef\arrowtime{\arrowtime}%
      }},
      mark=at position {#1-\arrowtime} with {\coordinate(@s);},
      mark=at position {#1} with {\coordinate(@e);
        \draw[-{Stealth[length=#2]}] (@s) -- (@e);}
    },
    postaction={decorate}
  }
}

\begin{document}
\title{Singlet-triplet superconductivity from repulsive interactions in a $p$-wave magnet}
\author{Hanbit Oh}
\thanks{Corresponding author:
\href{mailto:oh26@uwm.edu}{oh26@uwm.edu}}
\affiliation{Department of Physics, University of Wisconsin-Milwaukee, Milwaukee, Wisconsin 53201, USA}

\author{Laura Classen}
\affiliation{Physics Department, Technical University of Munich, TUM School of Natural Sciences, 85748 Garching, Germany}
\affiliation{Max-Planck-Institute for Solid State Research, 70569 Stuttgart, Germany}
\affiliation{Munich Center for Quantum Science and Technology (MCQST), Schellingstr. 4, D-80799 M\"unchen, Germany}

\author{Daniel F. Agterberg}
\thanks{Corresponding author: \href{mailto:agterber@uwm.edu}{agterber@uwm.edu}}
\affiliation{Department of Physics, University of Wisconsin-Milwaukee, Milwaukee, Wisconsin 53201, USA}

\date{\today}

\begin{abstract}
Motivated by recent proposals of $p$-wave magnets, which host odd-parity spin splitting, we study electrons near Van Hove singularities with local repulsive interactions.
We demonstrate how the symmetry of the $p$-wave spin splitting governs the competition among instabilities.
Using a parquet renormalization group (pRG) analysis of a simple one-sublattice model of a $p$-wave magnet, we find that superconductivity is the leading instability over the entire phase diagram for purely repulsive interactions.
The resulting superconducting state is generically a mixture of spin-singlet and spin-triplet pairing and approaches an equal-weight mixture when $T_c$ is small compared with the energy offset between each Van Hove electron and its inversion partner.
Extending the pRG analysis to a more realistic two-sublattice model, we find that superconductivity remains the leading instability, while the sublattice degree of freedom leads to a competition among various singlet-triplet pairing channels and even a finite-momentum equal-spin pair-density wave.
Our mechanism applies broadly to odd-parity magnets and opens a natural route to nodal topological superconductivity.
\end{abstract}

\maketitle

\textit{Introduction.---}Non-relativistic spin-splitting, the energy difference between spin-up and spin-down Bloch electrons, has recently become a focal point in quantum materials. While the emphasis has been on altermagnetism~\cite{Smejkal2022Beyond,Smejkal2022Emerging,Yuan2020Giant,naka:2019,hayami:2019,Roig2024Minimal}, in which this spin-splitting has even parity under sign change of the Bloch momentum, odd-parity spin-splittings have emerged as a new possibility~\cite{hellenes2024pwavemagnets,Hayami2020Spontaneous,Brekke2024Minimal,Yu2025Odd-Parity}. In $p$-wave magnets, a real space coplanar magnetic structure with a vanishing net magnetization gives rise to a $p$-wave collinear spin-splitting in momentum space~\cite{Song2025Electrical,yamada2025metallic,Zhou:2026}. Such a spin-splitting is of great interest since it can enable, for example, an Edelstein-like spin current response to electric field~\cite{edelstein1990spin,Yu2025Odd-Parity,Chakraborty2025Highly}.  

Importantly, in $p$-wave magnets, the origin of the spin-splitting is non-relativistic and originates from exchange interactions. 
Consequently, the corresponding energy scale can be much larger than that of spin-splittings generated by spin-orbit coupling. 
It is then natural to ask what electronic instabilities can occur in $p$-wave magnets. 
Previous studies of superconductivity (SC) in $p$-wave magnets have relied on mean-field theories with an attractive pairing interaction~\cite{Sun2025PseudoIsing,Khodas2026NonrelativisticIsing,kim2026topological,Luo2026Hidden,Froldi2026Superconducting,pal2026emergent}, leaving open whether electronic repulsion alone can drive superconductivity in these systems.
Here we first address this question by examining the well-known square-lattice Hubbard model with an added $p$-wave spin splitting.
In the $SU(2)$-symmetric Hubbard model, it has been known that the Van Hove singularities (VHSs) drive competing $d$-wave superconducting and spin-density-wave (SDW) states~\cite{dzyaloshinskii1987superconducting,Schulz1987Superconductivity,Furukawa1998Truncation}.
Using a parquet renormalization group (pRG) analysis~\cite{Nandkishore2012Chiral,2018IsobeUnconventional,Rao2025vanhovesingularites}, we find that $p$-wave spin splitting reshapes these VHS distribution and lifts this degeneracy, so that they drive only superconducting states through repulsive interactions. This is in sharp contrast to $d$-wave altermagnets (AM), where the even-parity spin splitting leads to density waves as the leading instabilities at the VHSs~\cite{Rao2025vanhovesingularites}.
We further find that these superconducting states generically form a mixture of spin-singlet and spin-triplet pairing, as in noncentrosymmetric superconductors~\cite{Gorkov2001Superconducting,Frigeri2004Superconductivity,Smidman2017Superconductivity}. Unlike the typically small admixture found in these systems, we often find an appreciable singlet-triplet mixture.

We further extend our analysis to a more realistic model in which nonsymmorphic symmetry is included since this is required to stabilize odd-parity antiferromagnetism~\cite{Yu2025Odd-Parity}.
The nonsymmorphic symmetry requires the inclusion of a sublattice with two degrees of freedom.
This model then has eight VHSs at three distinct energies, and to our knowledge it provides one of the few multi-sublattice settings in which a controlled weak-coupling pRG analysis has been carried out.
We find a similar phenomenology, with equal-weight singlet-triplet SC occupying most of the phase diagram.
However, the sublattice structure also stabilizes a finite-momentum equal-spin pair-density wave (PDW)~\cite{Agterberg2020Physics}.
This state is notable because it originates from repulsive interactions and 
does not require the presence of an external Zeeman field, unlike the Fulde-Ferrell-Larkin-Ovchinnikov state~\cite{fulde1964superconductivity,larkin1964nonuniform}. 
It thus provides a new example of a zero-field PDW state realized by VHSs in the weak-coupling limit~\cite{Wu2023Pair,Castro2023Emergence,Parthenios2025Spin}.

\begin{figure}[tb]
    \centering
    \includegraphics[width=\linewidth]
{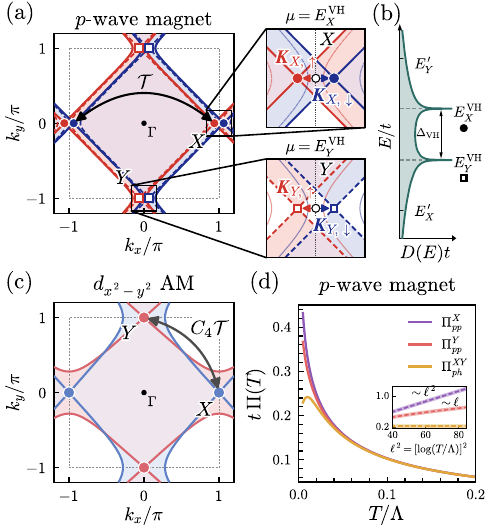}
    \vspace{-15pt}
    \caption{\textbf{Van Hove structure and bare susceptibilities of the $p$-wave magnet.}
(a) Spin-resolved energy contours of Eq.~\eqref{eq:h1} with $\lambda/t=0.4$. Red (blue) denotes $s=\uparrow$ ($\downarrow$). The solid and dashed curves are different energy level at $E_X^{\mathrm{VH}}$ and $E_Y^{\mathrm{VH}}$ respectively. Filled circles (open squares) mark the VHS points, $\bm K_{X,s}$ ($\bm K_{Y,s}$). $\lambda$ shifts the saddle points away from $X$ and $Y$ along $\mp\hat{x}$. The arrow connects the opposite-spin VHSs at the same energy related by $\mathcal{T}$, and the dispersion near these two are exactly the same.
Inset: zoom in near $X$ at $\mu=E_X^{\mathrm{VH}}$ and near $Y$ at $\mu=E_Y^{\mathrm{VH}}$.
(b) Total density of states, showing logarithmic singularities at $E_Y^{\mathrm{VH}}$ and $E_X^{\mathrm{VH}}$.
(c) For comparison, the $d_{x^2-y^2}$ altermagnet. Here the opposite-spin VHSs at the same energy are related by $C_4\mathcal{T}$, so their saddle curvatures are interchanged, which cuts off the $\log^2$ divergence of the opposite-spin Cooper channel.
(d) Energy dependence of the bare susceptibilities with leading $\log^2$ divergence at high T. 
We set $\lambda/t=0.3$ and $\mu=E_X^{\mathrm{VH}}$. Inset: low-temperature behavior $T\lesssim\Delta_{\mathrm VH}$ versus $\ell^2$. The dashed lines are fitting functions.
}
    \label{fig:1}
\end{figure}

\textit{Odd-parity spin splitting and VHSs.---}
We initially consider a simple square-lattice model with $p_x$-wave spin-splitting, 
\begin{equation}
H_{1}(\bk)
    =
    -2t(\cos k_x+\cos k_y)
    +\lambda f(\bm{k})\sigma_z -\mu ,
    \label{eq:h1}
\end{equation}
where $\sigma_z=\pm1$ labels the spin projection along the collinear axis.
$f(\bm{k})=\sin k_x$ encodes the odd-parity form factor and $\lambda$ is the strength of the spin splitting.
The subscript 1 in $H_1$ denotes the \textit{one-sublattice} model to distinguish from the two-sublattice model $H_2$ which will be discussed later. 
The simple model has a time-reversal symmetry $\mathcal{T}$ which enforces the property in its dispersion, $E_{s}(\bk)=E_{-s}(-\bk)$.
The odd parity spin-splitting has two key consequences on the VHS distribution.

First, the saddle points move away from the time-reversal-invariant momenta (TRIM) and become so-called type-II VHSs~\cite{Yao2015Topological}.
At $\lambda=0$ the spin-degenerate saddle points sit at
$X=(\pi,0)$ and $Y=(0,\pi)$. 
Because the gradient of an odd form factor is even in $\bk$, it need not vanish at $X$ and $Y$, and a finite $\lambda$ shifts the saddle points to
\begin{equation}
    \bm K_{X,s}=(\pi-s\kappa_X,0),
    \qquad
    \bm K_{Y,s}=(-s\kappa_Y,\pi),
    \label{eq:VHS_positions}
\end{equation}
with $\kappa_{X,Y}=\tan^{-1}(\lambda/2t)$
[Fig.~\ref{fig:1}(a)].  
This is in sharp contrast to even-parity $d$-wave AM, where the VHSs remain pinned at TRIM points, $X$ and $Y$ [Fig.~\ref{fig:1}(c)]~\cite{Rao2025vanhovesingularites}.  
The shift is accompanied by a splitting of the Van Hove (VH) energies of the two families,
$\Delta_{\mathrm{VH}}\equiv E^{\mathrm{VH}}_{X}-E^{\mathrm{VH}}_{Y}
=2\sqrt{4t^2+\lambda^2}-4t\simeq\lambda^2/2t$.
Expanding around the VH points, $\bm{q}=\bm{k}-\bm{K}_{\alpha,s}$, and measuring energies from $E^{\mathrm{VH}}_{X}$, the dispersions read
\begin{align}
\epsilon_{X,s}(\bm q)&=-a\, q_x^2 + b\, q_y^2 + \mathcal{O}(q^4), 
\nonumber\\
\epsilon_{Y,s}(\bm q)&=-\Delta_{\mathrm{VH}} + a\, q_x^2 - b\, q_y^2 + \mathcal{O}(q^4),
\label{eq:vhs_dispersion}
\end{align}
with $a=\sqrt{4t^2+\lambda^2}/2$ and $b=t$.
Both dispersions are independent of the spin $s$ and even in $\bm q$.
The density of states shows two logarithmic peaks [Fig.~\ref{fig:1}(b)].
Note that throughout this work, we set the chemical potential at the upper one, $\mu=E^{\mathrm{VH}}_{X}$.

Second, $\mathcal{T}$ symmetry ties the two opposite-spin saddle points of a given family to each other: $\bm K_{\alpha,\up}=-\bm K_{\alpha,\down}$ ($\alpha=X,Y$), and the two have identical dispersions [Eq.~\eqref{eq:vhs_dispersion}]. 
Therefore opposite-spin Cooper pairs with zero momentum can be formed by these VHS electrons [Fig.~\ref{fig:1}(a)], and the corresponding particle-particle (pp) bubble retains its full $\log^2$ divergence for any $\lambda$.
This is another essential difference from the $d$-wave AM. 
There, opposite-spin VHSs at the same energy are either related by $C_4\mathcal{T}$, which interchanges their curvatures, or, for $d_{xy}$ splitting, located at the same momentum but with principal axes tilted in opposite directions [Fig.~\ref{fig:1}(c)].
In either case their dispersions do not coincide, so the opposite-spin pp channel loses its full $\log^2$ divergence~\cite{Rao2025vanhovesingularites}.
In the $p$-wave magnet, it is instead the particle-hole (ph) channel that is cut off, because density waves connect $X$ and $Y$, whose VHS energies differ by $\Delta_\mathrm{VH}$.
The roles of the two channels are thus reversed compared to the $d$-wave AM.

These features can be directly seen in  bare susceptibility calculations. We focus on the intrapatch pp bubbles, $\Pi^{X}_{pp}$ and $\Pi^{Y}_{pp}$, and the interpatch ph bubble $\Pi^{XY}_{ph}$ which have the leading divergence for $T\gtrsim\Delta_{\mathrm VH}$. Since the patch dispersions in Eq.~\eqref{eq:vhs_dispersion} do not depend on spin, each bubble takes the same value for equal and opposite spins, and both spin configurations are included on an equal footing.
Above the splitting, $T\gg\Delta_{\mathrm{VH}}$, the $X$ and $Y$ patches are indistinguishable and all three bubbles grow as $\ell^{2}$, with $\ell=\log(\Lambda/T)$, as in the $\lambda=0$ model. 
All other one-loop bubbles having a single log, such as the intrapatch ph and the interpatch pp, grow at most as $\ell$ and are neglected in this regime. 
Below $\Delta_{\mathrm{VH}}$ they separate. 
In this regime, the pp bubble at $X$ still diverges as $\ell^2$, whereas that at $Y$ grows only as $\ell\log(\Lambda/\Delta_{\mathrm{VH}})$, i.e. %and grows
with the same power of $\ell$ as the neglected single-log bubbles. The interpatch ph bubble even saturates below $\Delta_{\mathrm{VH}}$. 
Fig.~\ref{fig:3}(d) shows the numerically evaluated bubbles $\Pi^{X}_{pp}$, $\Pi^{Y}_{pp}$, and $\Pi^{XY}_{ph}$ as functions of temperature. See Supplemental Materials (SM) Sec.~\ref{sec:1sub_model} for details.

\textit{Parquet RG and competing orders.---}
We next perform the pRG for the electrons in the four patches around $\bm K_{\alpha,s}$ to determine how interactions lead the competition between these channels~\cite{dzyaloshinskii1987superconducting,Schulz1987Superconductivity,Nandkishore2012Chiral,2018IsobeUnconventional,Rao2025vanhovesingularites}.
As in the standard RG, we consider generalizations of the four processes of Fig.~\ref{fig:2}(a): exchange ($g_1$), forward ($g_2$), pair-hopping ($g_3$) and intrapatch ($g_4$) scattering.
The magnetism reduces the spin symmetry to rotations about the magnetic axis, so the parallel- and antiparallel-spin process become independent couplings, $g_{pi}$ and $g_{ai}$
[Figs.~\ref{fig:2}(b,c)].
In addition, $X$ and $Y$ are no longer related
by $C_4$, so the intrapatch couplings split into $g^{X}_{4}$ and $g^{Y}_{4}$. This gives 10 independent coupling constants in total and their explicit forms are given in Eqs.(\ref{eq:EM_HP},\ref{eq:EM_HA}).

\begin{figure}[tb]
    \centering
    \includegraphics[width=\linewidth]
{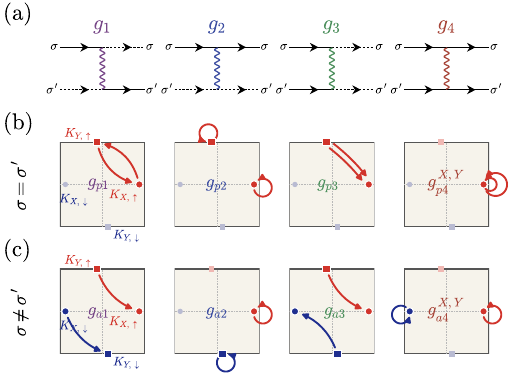}
    \caption{\textbf{Patch interactions.}
    (a) The four scattering processes between the Van Hove patches.
    Solid (dashed) lines denote electrons in the $X$ ($Y$) patches.
    (b),(c) Representative momentum-space processes for parallel
    ($\sigma=\sigma'$) and antiparallel ($\sigma\neq\sigma'$) spins.
    Each arrow shows the scattering of one electron and is colored by
    its spin.
    }
    \label{fig:2}
\end{figure}

We take $y\equiv-\Pi^{X}_{pp}\propto\log^{2}(\Lambda/T)$ as the running scale and encode the other two bubbles in $d(y)=-d\Pi^{Y}_{pp}/dy$ and $\tilde d(y)=d\Pi^{XY}_{ph}/dy$, which we evaluate numerically from Eq.~\eqref{eq:h1}.
Note that retaining only these three bubbles is controlled when $T\gg\Delta_{\mathrm{VH}}$, 
while we check the effect of the additional single-logarithmic channels for $T\lesssim\Delta_{\mathrm{VH}}$. This is provided in SM for completeness~\cite{SM}.
The resulting one-loop flow equations, $\dot{g}_i=dg_i/dy$, are quadratic in $g_i$ with coefficients set by $d$ and $\tilde d$ [Eq.~\eqref{eq:RG_flow}]. 
For $\lambda=0$, $d=\tilde d=1$ and they reduce to the RG results of conventional square-lattice Hubbard model at VH filling~\cite{Furukawa1998Truncation}. 
For $\lambda\neq0$, both functions decrease, with $\tilde d$ decreasing faster than $d$.
The couplings diverge at a critical scale $y_c$ as $g_i\simeq G_i/(y_c-y)$, and the susceptibility of a test vertex in channel $j$ behaves as $\chi_j\propto(y_c-y)^{1-2\alpha_j}$, with the exponents $\alpha_j$.
An instability requires $\alpha_j>1/2$, and the largest exponent identifies the leading order.
The candidate orders are: charge-density wave (CDW) and longitudinal spin-density wave (SDW-$l$) with commensurate wave vector $(\pi,\pi)$, which connect equal-spin saddle points; transverse spin-density wave (SDW-$t$) with incommensurate wave vectors $\pm(\pi-2\kappa,\pi)$, which connect opposite-spin saddle points; and opposite-spin superconductivity with zero center-of-mass momentum. Here, SC$_\pm$ denote the two eigenmodes of the coupled superconducting vertex $(\Delta_X,\Delta_Y)$ [Eq.~\eqref{eq:SC_vertex_flow}], with SC$_+$ (SC$_-$) corresponding to the larger (smaller) eigenvalue, i.e., the larger (smaller) exponent $\alpha_{\rm SC}$.
Note that, even though VHSs are off from TRIM points, equal-spin pairing is still absent at zero momentum because $E_s(\bm k)\neq E_s(-\bm k)$, so the inversion partner of the given VH electron lies away from the VH energy.

As bare interactions, we take an on-site repulsion $U$ and a nearest-neighbor repulsion $V$ with $V(\bm q)=2V(\cos q_x+\cos q_y)$.
Initial conditions are then set by \eqref{eq:initial_condition}. 
Throughout this work, we set $U=0.1t$. 
The resulting phase diagram is shown in Fig.~\ref{fig:3}(a). 
Over the whole range of $V/U$ and $\lambda/t$ considered, the leading instability is the superconducting state except on the fine-tuned line $V/U=1/4$.
The color scale shows the exponent difference $\alpha_{\mathrm{SC}}-\alpha_{\mathrm{DW}}$. Its positive value means that superconductivity wins over density waves, and the difference grows monotonically with $\lambda$.
The instability occurs above $\Delta_{\mathrm{VH}}$, $T_c>\Delta_{\mathrm{VH}}$, for small $\lambda$ and below it, $T_c<\Delta_{\mathrm{VH}}$ for large $\lambda$. We denote by $\lambda^*$ the value satifying $T_c=\Delta_{\mathrm{VH}}$, shown as the red line in Fig.~\ref{fig:3}(a). For $\lambda<\lambda^*$, the selection of $\log^2$ channels is controlled and the single-log channels can be neglected. For $\lambda>\lambda^*$, these channels become comparable to those retained. Including these explicitly, we find that SC$_+$ is still the leading instability [Fig.~\ref{fig:S_singlelog}].
\begin{figure}[tb]
    \centering
    \includegraphics[width=\linewidth]
{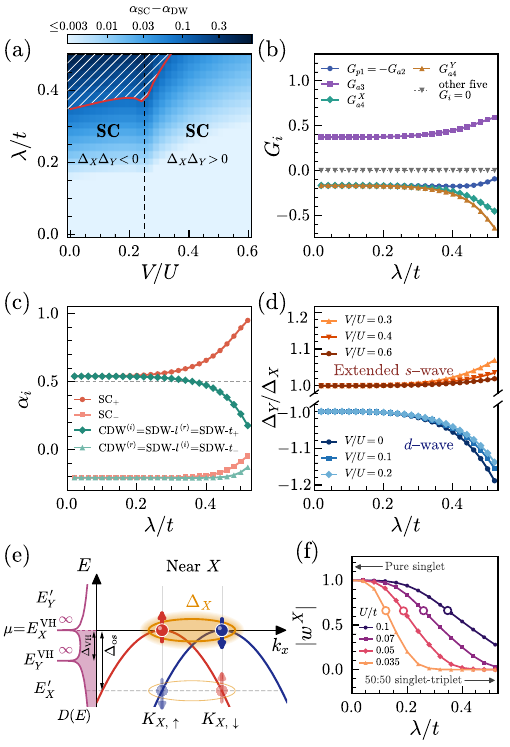}
\caption{\textbf{Parquet RG results of the one-sublattice model Eq.~\eqref{eq:h1}.}
We set $\mu=E_X^{\mathrm{VH}}$ and $U=0.1t$ unless stated otherwise.
(a) $\alpha_{\mathrm{SC}}-\alpha_{\mathrm{DW}}$ in the $(V/U,\lambda/t)$ plane. $\mathrm{SC}_{+}$ is leading everywhere except on the fine-tuned line $V/U=1/4$ (dashed). 
This line separates $d$-wave-like ($\Delta_X\Delta_Y<0$) from extended $s$-wave-like ($\Delta_X\Delta_Y>0$) SC$_+$.
The red line marks $T_c=\Delta_{\mathrm{VH}}$, and the dashed region above it has $T_c<\Delta_{\mathrm{VH}}$. 
(b) Fixed-trajectory coefficients $G_i$ and (c) vertex exponents $\alpha_i$ at $V/U=0$. The dashed line in (c) marks the threshold $\alpha=1/2$. For all $\lambda$, $\mathrm{SC}_{+}$ has the largest exponent.
(d) Gap ratio $\Delta_Y/\Delta_X$ of the $\mathrm{SC}_{+}$ eigenvector versus $\lambda$ for several $V/U$.
(e) Schematics of bands near $X$ at $\mu=E^\mathrm{VH}_X$. The pair $P_{\downarrow\uparrow}$ at the Van Hove points (filled) is paired by $\Delta_X$, while its opposite-spin partner $P_{\uparrow\downarrow}$ (faded) lies $\Delta_{\mathrm{os}}$ below $\mu$. (f) Relative amplitude $|w^X|$ [Eq.~\eqref{eq:wX}] versus $\lambda$ for $V=0$ and several $U$, estimated from $T_c$ with $|\Delta_X|\simeq1.76\,T_c$. The state evolves from pure-singlet pairing ($|w^X|=1$) to 50:50 singlet--triplet pairing ($|w^X|=0$).
Open circles mark $\lambda^*$, where $T_c=\Delta_{\mathrm{VH}}$, for each $U$.}
    \label{fig:3}
\end{figure}

To see why the odd-parity splitting favors superconductivity, we examine fixed trajectories. In particular, in Figs.~\ref{fig:3}(b,c), we illustrate the case with only
Hubbard interaction $V=0$. Other values of $V/U$ behave qualitatively the same.
On this trajectory, only the antiparallel pair hopping $G_{a3}>0$, the intrapatch couplings $G^{X,Y}_{a4}<0$, and $G_{a2}=-G_{p1}>0$ remain finite. At $\lambda=0$, this reproduces the well known limit where $d$-wave pairing and $(\pi,\pi)$ SDW diverge with the same exponent, $\alpha\simeq0.54$~\cite{Furukawa1998Truncation}. 

Therefore, for small $\lambda$, it is hard to distinguish between these because the two exponents remain close. 
A finite $\lambda$ lifts this degeneracy, because $\tilde d,d$ are suppressed, while the pp kernel at $X$ is not, only the pair-hopping ($g_{3a}$) and intrapatch couplings ($g_{4a}$) continue to
grow [Eq.~\eqref{eq:RG_flow}]. Consequently, $\alpha_{\mathrm{SC},+}$ increases with $\lambda$ towards one. At the same time, the density-wave exponent decreases and even falls below $1/2$ at $\lambda\simeq0.36t$.

The only exception is the line $V/U=1/4$ [Fig.~\ref{fig:3}(a)]. There $g_{a1}(0)=g_{a3}(0)=U-4V=0$, and since $\dot g_{a1}\propto g_{a1}$ and $\dot g_{a3}\propto g_{a3}$, both remain zero. Then the flow can be solved analytically, giving $G_{a2}=-G_{p1}=1/\tilde d_c$ with all other $G_i=0$ (See SM for details).
Then due to the absence of $G_{a3}$, $G_{a4}$ [Eq.~\eqref{eq:SC_vertex_flow}], superconductivity is never generated within the trajectory, while the six density-wave channels diverge together with $\alpha=1$. However, this trajectory is unstable, thus a small perturbation from $V/U=1/4$ restores $\mathrm{SC}_{+}$.

\textit{Superconducting instability.---}
A central property of SC$_+$ is that it mixes spin-singlet and spin-triplet pairing. The superconducting vertex pairs the two opposite-spin saddle points of each family, $H_{\mathrm{SC}}=\sum_{\bm{k}}\sum_{\alpha=X,Y}\Delta_\alpha\psi^\dagger_{\alpha,\downarrow}(-\bm k )\psi^\dagger_{\alpha,\uparrow}(\bm{k})+\mathrm{H.c.}$. 
Using the notation, $\psi^\dagger_{\alpha,\uparrow}(\bm{k})=c_{\sigma} (\bm{K}_{\alpha,\sigma}+\bm{k})$, let us focus $\Delta_X$ and consider the electrons at $\bm K_{X,\uparrow}$ and $\bm K_{X,\down}$. 
First, the pair $P_{\downarrow \uparrow}
=
c^\dagger_{\downarrow}(\bm{K}_{X,\downarrow})c^\dagger_{\uparrow}(\bm{K}_{X,\uparrow})=
\psi^\dagger_{X,\downarrow}(0)\psi^\dagger_{X,\uparrow}(0)$ is formed by two VHS electrons. 
Meanwhile the opposite-spin partner $P_{\uparrow\downarrow}=c^\dagger_{\uparrow}(\bm{K}_{X,\downarrow})c^\dagger_{\downarrow}(\bm{K}_{X,\uparrow})=\psi^\dagger_{\uparrow}(2\kappa)
\psi_{\downarrow}^\dagger(-2\kappa)$ are at the same momenta, but these lie away from the VHSs by $\Delta_{\mathrm{os}}\equiv|E^{\mathrm{VH}}_{X}-E'_{X}|$. Here, $E'_{X}=E_{-s}(\bm K_{X,s})$ is the energy of the opposite-spin state, which gives $\Delta_{\mathrm{os}}=2\lambda^2/\sqrt{4t^2+\lambda^2}\simeq2\Delta_{\mathrm{VH}}$.
Crucially, since the momentum-space separation between the opposite-spin VH points, $\bm{K}_{X,\downarrow}-\bm{K}_{X,\uparrow}\equiv 2\kappa\simeq\lambda/t$, is much less than the patch size $\Lambda$, both $P_{\downarrow\uparrow}$ and $P_{\uparrow\downarrow}$ lie within the patches and are paired by the same $\Delta_X$.

Then the relative amplitude of $\langle P^X_{\downarrow\uparrow}\rangle$ and $\langle P^X_{\uparrow\downarrow}\rangle$ is set by the energies of pairs measured from the chemical potential, $\xi_{\bm{k}}=E_{\bm{k}}-\mu$, and the pairing gap $|\Delta_X|$,
\begin{equation}
  w^X\equiv-\frac{\langle P^X_{\uparrow\downarrow}\rangle}{\langle P^X_{\downarrow\uparrow}\rangle}
\simeq\frac{|\Delta_X|}{\sqrt{|\Delta_X|^2+\Delta_{\mathrm{os}}^2}}.
\label{eq:wX}
\end{equation}
This is achieved by $ w^X\simeq 
\frac{\sqrt{|\Delta_X|^2+\xi_{\downarrow\uparrow}^2}}{\sqrt{|\Delta_X|^2+\xi_{\uparrow \downarrow}^2}}$ with substituting $\xi_{\downarrow\uparrow}=0$ and $\xi_{\uparrow \downarrow}=\Delta_\mathrm{os}$.
Then the value $0\le |w^X|\le1$ controls the mixing between singlet $S=(P^X_{\downarrow\uparrow}-P^X_{\uparrow\downarrow})/\sqrt{2}$ and $S_z=0$ triplet $T_0=(P^X_{\downarrow\uparrow}+P^X_{\uparrow\downarrow})/\sqrt{2}$.
In the small-$\lambda$ limit, where $\Delta_{\mathrm{os}}\ll|\Delta_X|\sim T_c$, both partners are active with equal amplitude, $w^X\to1$. The pairing is then $P_{\downarrow\uparrow}-P_{\uparrow\downarrow}\propto S$, a pure singlet. 
In the opposite limit, $\Delta_{\mathrm{os}}\gg|\Delta_X|\sim T_c$, $P_{\uparrow\downarrow}$ is suppressed, $w^X\to0$, and only $P_{\downarrow\uparrow}=(S+T_0)/\sqrt{2}$ remains. This is an \emph{equal-weight singlet-triplet mixture}.
The state therefore evolves continuously from singlet-dominated pairing to equal-weight singlet--triplet pairing as $T_c/\Delta_{\mathrm{os}}$ decreases or $\lambda$ increases. In Fig.~\ref{fig:3}(f), we illustrate the evolution.

The $Y$-component $\Delta_Y$ behaves differently. At the $Y$ VH points, the pair $P_{\downarrow\uparrow}$ lies below the Fermi level, $\xi_{\downarrow\uparrow}=-\Delta_{\mathrm{VH}}$, while its opposite-spin partner lies above it by the same amount, $\xi_{\uparrow\downarrow}=\Delta_{\mathrm{os}}-\Delta_{\mathrm{VH}}\simeq\Delta_{\mathrm{VH}}$. Since the two partners are almost symmetric about the Fermi level [Fig.~\ref{fig:1}(b)], $w^Y\simeq1$ and the $Y$ component remains nearly a pure singlet. The equal-weight structure is therefore carried by the $X$-patch pairing alone.
Mean-field studies of $p$-wave magnets, which address the physics away from the VHSs, find mixed singlet--triplet states with comparable weights of the two components~\cite{Sun2025PseudoIsing,kim2026topological,Luo2026Hidden}. Here the mixing is instead controlled by $T_c/\Delta_{\mathrm{os}}$, since the pairing scale generated by VHSs can lie well below the spin splitting, so the state approaches the equal-weight limit.

Finally, we turn to the momentum structure of $\mathrm{SC}_{+}$. The two amplitudes $(\Delta_X,\Delta_Y)$ are coupled by pair hopping [Eq.~\eqref{eq:SC_vertex_flow}]. 
On the fixed trajectory, the eigenmode with the larger exponent
$\alpha_{\mathrm{SC},+}$ satisfies 
\begin{equation}
    \Delta_Y/\Delta_X=-\,\mathrm{sgn}(G_{a3})\,d_c^{-1/2},
    \label{eq:gap_ratio}
\end{equation}
with $d_c\equiv d(y_c)$. 
For $V/U<1/4$, we have $G_{a3}>0$,
and the gap changes sign between the two families. For $V/U>1/4$, $G_{a3}<0$ and the gap keeps its sign [Fig.~\ref{fig:3}(f)].
At $\lambda=0$, $C_4$ symmetry enforces $|\Delta_Y/\Delta_X|=1$. A finite $\lambda$ breaks this symmetry, so the $s$- and $d$-wave components no longer belong to different representations, and $\mathrm{SC}_{+}$ becomes an $s+d$ mixture. 
Thus we call these gaps $d$-wave-like ($\Delta_X\Delta_Y<0$) and extended $s$-wave-like ($\Delta_X\Delta_Y>0$).
Note that at $V=0$ and $\lambda\to0$, we have $\Delta_Y/\Delta_X=-1$, and this $d$-wave-like state corresponds to $d_{x^2-y^2}$ pairing, reproducing the result of the $SU(2)$-symmetric Hubbard model~\cite{Schulz1987Superconductivity,Furukawa1998Truncation}.

\begin{figure}
    \centering
    \includegraphics[width=\linewidth]{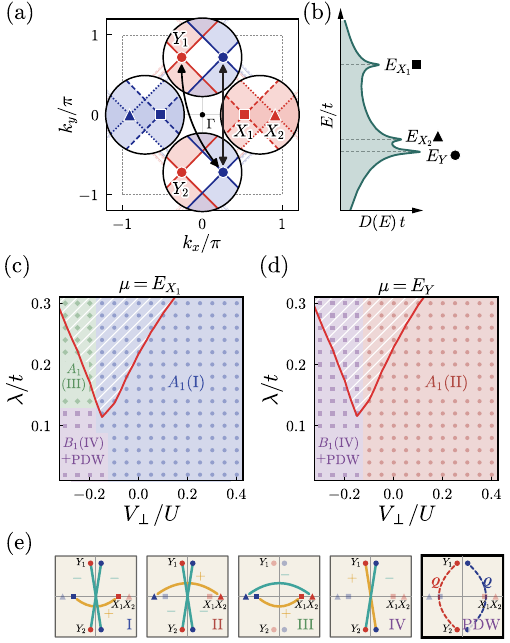}
\caption{\textbf{Parquet RG results of the two-sublattice model, Eq.~\eqref{eq:h2}.} We set $\lambda/t=0.4$, $t_\perp=0.3t$. 
(a) Eight van Hove points of the four bands. The insets show the regions near $\pm X$ and $\pm Y$, each showing two VHSs with their constant-energy contours at the respective VH energy and the occupied regions (shaded). Red (blue) denotes spin up (down). Circles, squares, and triangles mark the VHSs at $E_Y$, $E_{X_1}$, and $E_{X_2}$. The arrows indicate the zero-momentum opposite-spin pair between $Y_1$ and its $\mathcal{T}$ partner ($\bm{Q}=0$) and the finite-momentum equal-spin pair between $Y_1$ and $Y_2$ ($\bm{Q}\neq0$).
(b) Total density of states, showing VH peaks at $E_Y<E_{X_2}<E_{X_1}$.
(c),(d) Leading instability in the $(V_\perp/U,\lambda/t)$ plane for (c) $\mu=E_{X_1}$ and (d) $\mu=E_Y$ and $V/U=0.1$. The leading instability is always SC or, in purple region, an equal-spin PDW degenerate with SC IV.
The red line marks $\lambda^*$ satisfying $T_c=E_{X_1}-E_Y$, and the hatched region has $T_c<E_{X_1}-E_Y$.
(e) Gap structure of SC and PDW.
I-IV panels illustrate SC states where solid links and $\pm$ show the pairing amplitudes and their signs. The last panel is for the equal-spin PDW $\langle\psi_{Y_1\sigma}\psi_{Y_2\sigma}\rangle$ with momentum $\bm{Q}$, which diverges together with SC IV.
}
    \label{fig:4}
\end{figure}

\textit{Robustness in a two-sublattice model.---}
A microscopic $p$-wave magnet requires a nonsymmorphic symmetry and hence is realized in at least a $4\times4$ (sublattice $\times$ spin) model. 
We consider the effective $p$-wave magnet model, derived from the full eight-band model~\cite{Yu2025Odd-Parity}. 
This effective model can be justified when the low energy states lie near the $X,Y$ points. 
This has been checked by perturbation theory and by comparison with the full eight-band model. This effective model is
\begin{equation}
H_{2}(\bk)=\epsilon_0(\bm k)\,\tau_0\sigma_0+\epsilon_\perp(\bm k)\,\tau_x\sigma_0+\lambda\sin k_x\,\tau_0\sigma_z ,
\label{eq:h2}
\end{equation}
where $\epsilon_0(\bm k)=-2t(\cos k_x+\cos k_y)-\mu$, $\epsilon_\perp(\bm k)=2t_\perp\cos(k_x/2)\cos(k_y/2)$ is the intersublattice hopping, and $\tau_i$ act on the sublattice index. The bands are $E_{\eta s}(\bm k)=\epsilon_0+\eta\,\epsilon_\perp+s\lambda\sin k_x$ with $\eta=\pm$. 
The intersublattice hopping splits the $X$ saddle points of each spin into two VHSs, $X_1$ and $X_2$, at different energies. At $Y$, by contrast, $\epsilon_\perp$ vanishes on the zone boundary $k_y=\pi$, and the two saddle points $Y_1$ and $Y_2$ remain degenerate. 
Here, we label $\{X_i, Y_i\}\equiv \{X_{i,\uparrow},Y_{i,\uparrow}\}$ as the VHSs of spin-up state. The spin-down partner VHSs are at opposite momentum, $ \{X_{i,\downarrow},Y_{i,\downarrow}\}= \{-X_i, -Y_i\}$ with $i=1,2$.
In total, there are eight VHSs (including both spin up/down) that lie at three different energies, $E_Y<E_{X_2}<E_{X_1}$. The Fermi surfaces at each Van Hove filling are shown in Fig.~\ref{fig:4}(a,b) and Fig.~\ref{fig:s_2sub_FS}.

Like $H_1$, $H_2$ is invariant under  time-reversal $\Theta=i\sigma_y\mathcal K$, which in the microscopic magnet corresponds to $\mathcal T\tau$ or $\mathcal T C_2^\perp$ rather than to the physical time reversal broken by the magnetic order~\cite{Luo2026Hidden}. Both can enforce
$E_{s,\eta}(\bk)=E_{-s,\eta}(-\bk)$, and hence $E_{s,\eta}(\bm K+\bm q)=E_{-s,\eta}(-\bm K-\bm q)$ around every VHSs. Zero-momentum opposite-spin pairs in all families $X_i$ and $Y_i$ ($i=1,2$) therefore keep the Cooper logarithm and thus can have the full $\log^2$ divergence of the pp bubble once the chemical potential lies at their VHS energy.
A new feature here is that we now have four VHSs, $Y_{1,2}$ accounting for both spins, lie at the same energy $\mu=E_Y$. In this case, in addition to the $\bm Q=0$ pair, ($Y_{1}\!\uparrow$, $-Y_{1}\!\downarrow$),
the finite-momentum pair ($Y_{1}\!\uparrow$, $Y_{2}\!\uparrow$) should be considered (see Fig.~\ref{fig:4}), since they are also $\log^2$ divergent [Fig.~\ref{fig:s_2sub_bubble}(b)]. 
This is thus different from the one-sublattice model, where we cannot consider any possible same-spin pairing.

Including all eight patches gives seven independent bubbles that grow as $\log^2$ at high-temperature and 22 independent couplings.
Following the same procedure, we solve their flow equations and instabilities [Eq.~\eqref{eq:RG_2sub}]. 
Here, we take as bare interactions the on-site $U$, an intrasublattice nearest-neighbor repulsion $V$, and an intersublattice repulsion $V_\perp$, all projected onto the band basis. We fix $t_\perp=0.3t$ and $V/U=0.1$ with $U=0.1t$, and place $\mu$ either at $E_{X_1}$ or at $E_Y$.  
The case $\mu=E_{X_2}$, which lies close to $E_Y$, gives almost the same phase diagram as $\mu=E_Y$ (Fig.~\ref{fig:S_muEX2} in SM).
Our main conclusion from the two sublattice case is as follows: (i) Equal-weight singlet--triplet SC remains the leading instability as in the one-sublattice model, and (ii) the richer VH structure yields a richer variety of pairing states with a competing PDW~[Fig.~\ref{fig:4}].

We classify the superconducting states by the irreps of $C_{2v}=\{E,C_2(x),M_y,M_z\}$, the point group of the model [Table~\ref{tab:SC_2sub}]. The $M_y$-even states with $\Delta_{Y_1}=\Delta_{Y_2}$ belong to $A_1$, and the $M_y$-odd state with $\Delta_{Y_1}=-\Delta_{Y_2}$ belongs to $B_1$. Within $A_1$, we further label the states by their gap structure. State I (II) has its gap on $X_1$ ($X_2$) and $Y_{1,2}$ with $\Delta_X\Delta_Y<0$. State III has its gap on $X_{1,2}$ with $\Delta_{X_1}=-\Delta_{X_2}$. The $B_1$ state, state IV, has $\Delta_{Y_1}=-\Delta_{Y_2}$ and $\Delta_{X_{1,2}}=0$. 
The spin structure is again set by the energy offset of the opposite-spin partner at the same momenta.
The corresponding pairing approaches an equal-weight singlet-triplet mixture at large $\lambda$ regime.

Fig.~\ref{fig:4}(c,d) show the phase diagrams in the $(V_\perp/U,\lambda/t)$ plane.  Over the entire plane, the leading instability is either superconductivity or a PDW, and the density-wave exponents stay below $1/2$ ($\alpha_{\mathrm{DW}}\lesssim0.43$). For $V_\perp>0$, the leading state is I at $\mu=E_{X_1}$ and II at $\mu=E_Y$. An $V_\perp<0$ instead favors III at $\mu=E_{X_1}$ and IV at $\mu=E_Y$. Since I, II, and III share the same irrep $A_1$, the boundaries among these three are crossovers rather than phase transitions, and only state IV is distinct by symmetry. Moreover, state IV is degenerate with an equal-spin PDW $\langle\psi_{Y_1\sigma}\psi_{Y_2\sigma}\rangle$ with momentum $Q=Y_1+Y_2 \approx (-\frac{\lambda}{t},0)$ [Fig.~\ref{fig:4}(e)]. We find that the two order parameters obey identical flow equations with identical bare couplings and therefore share the same exponent, $\alpha_{\mathrm{IV}}=\alpha_{\mathrm{PDW}}$ (See SM Sec~\ref{sec:2sub_testvertices} for details). Thus, the degeneracy is accidental and will be removed upon inclusion of other bare interactions.  The sublattice degree of freedom thus leads to a competition among several equal-weight singlet--triplet superconductors and even a finite-momentum equal-spin state.
A similar equal-spin PDW in $p$-wave magnets has been studied in an extended attractive Hubbard model away from VHSs within mean-field theory~\cite{Lee2026Equal-spin}.
Here, by contrast, it emerges from the VHSs at weak repulsive coupling, which also distinguishes it from PDWs in the strong-coupling regime of correlated models~\cite{Berg2010Pair-Density-Wave,Zhang2022Pair-density-wave,Jiang2024Pair-Density-Wave,Oh2026Incommensurate}.

We remark that the restriction to $\log^2$ channels in the pRG of the two-sublattice model is controlled for $T_c>E_{X_1}-E_Y$, which defines $\lambda^*(V_\perp)$, e.g., $\lambda^*\simeq0.22t$ at $V_\perp=0$.
Its $V_\perp$ dependence is shown in Fig.~\ref{fig:4}(c,d) with red line. 
Although we have not included all single-log channels, the pp channel at the VHS on the Fermi level keeps its full $\log^2$ divergence even for $T_c<E_{X_1}-E_Y$, and we therefore expect the competition tendency to remain unchanged.

\textit{Discussion and conclusion.}---We have performed a pRG analysis of a VHS model of a $p$-wave magnet and have shown that repulsive interactions drive superconductivity with mixed spin-singlet -triplet pairing, which evolves from an $d$-wave-like to an extended $s$-wave-like gap across changing the interaction ratio $V/U$. 
We further showed that this conclusion is universal based on the leading $\log^2$ behavior of pp corrections and even holds in the two-sublattice model.
In a two-sublattice model, superconductivity remains leading, with several competing singlet-triplet states and even an equal-spin PDW appearing at weak coupling.

Although our study focuses on a specific square-lattice $p$-wave magnet model, it is straightforward to extend the singlet--triplet SC mechanism to any system in which (i) opposite-spin VHSs sit at opposite momenta $\pm\bm K$ with $\epsilon_{\alpha,s}(\bm q)=\epsilon_{\alpha,-s}(-\bm q)$, and (ii) their inversion partners lie away from the Van Hove energy.
For example, in the $f$-wave magnet on the triangular lattice~\cite{Luo2026Hidden}, the odd-parity spin splitting shifts the three VHSs at $M_{i}$ ($i=1,2,3$) to six spin-resolved VHSs at $M_{i}+s\,\delta\bm q_{i}$ with $s=\pm1$. This setup still satisfies both conditions, and we expect that repulsive interactions can likewise induce an equal-weight singlet--triplet SC.
The effect of the spin splitting is, however, less dramatic than on the square lattice.
For three VHSs at $M_i$, as on the triangular or honeycomb lattices, a chiral superconductivity already dominates over density waves even in the $SU(2)$-symmetric perfect nesting case~\cite{Nandkishore2012Chiral}.
The odd-parity splitting is thus expected to preserve this preference but turn the pairing into a singlet-triplet mixture.

Finally, we comment on the topology of the mixed singlet--triplet SC. Previous works showed that such a state is a natural candidate for nodal topological superconductivity, classifying it to class DIII~\cite{kim2026topological} or BDI~\cite{Luo2026Hidden}. The difference stems from the choice of the model and the corresponding effective time-reversal operator. The effective one-sublattice model has the $\Theta=i\sigma_yK$ with $\Theta^2=-1$~\cite{kim2026topological}, whereas the microscopic model instead has $\tilde{\mathcal T}=\mathcal T C_2^\perp$ with $\tilde{\mathcal T}^2=+1$~\cite{Luo2026Hidden}. The two can be reconciled by considering a momentum space nodal classification, which includes the symmetry keep momentum invariant. 
In AZ+$\mathcal{I}$ classification~\cite{Bzdusek2017Robust}, both have the same chiral symmetry, where the nodes belong to class AIII. Thus, point nodes in $d=2$ (line nodes in $d=3$) are protected by a $\mathbb Z$ winding number and give rise to Majorana flat bands at the edges~\cite{Sato2011Topology,Schnyder2011Topological}.
In the $s^{+-}
$ state in our case, the sign change guarantees the existence of nodes and will thus be topological.

\textit{Acknowledgements.---}
H.O. and D.F.A. are supported by the U.S. Department of Energy, Office of Basic Energy Sciences, Division of Materials Sciences and Engineering under Award No. DE-SC0021971 and Simons Foundation Grant No. SFI-MPS-NFS-00006741-02. 
L.C. was supported by the Simons Foundation SFI-MPS-NFS-00006741-11, the Deutsche Forschungsgemeinschaft (DFG, German Research Foundation) under Germany’s Excellence Strategy–EXC–2111–390814868 and TRR 360 – 492547816, and by the European Union (ERC-2023-STG, Project 101115758 - QuantEmerge). Views and opinions expressed are, however, those of the authors only and do not necessarily reflect those of the European Union or the European Research Council Executive Agency. Neither the European Union nor the granting authority can be held responsible for them.
Claude Opus 5.5 were used to assist in improving language and figures.

%\bibliography{ref_p-wave}
%apsrev4-2.bst 2019-01-14 (MD) hand-edited version of apsrev4-1.bst
%Control: key (0)
%Control: author (8) initials jnrlst
%Control: editor formatted (1) identically to author
%Control: production of article title (0) allowed
%Control: page (0) single
%Control: year (1) truncated
%Control: production of eprint (0) enabled
%

\begin{center}
\vspace{10pt}
\textbf{End Matter}
\end{center} 
\textit{Low-energy effective interactions.}---We keep only fermions near the four spin-resolved saddle points $\bm K_{\alpha,s}$ with $\alpha=X,Y$. There are 10 symmetry-allowed interactions splitting into parallel- and antiparallel-spin parts,
\begin{align}
H_{\mathrm{int},p}=\;&\frac{1}{2}\sum_{\sigma}\sum_{\alpha\neq\beta}\Big(g_{p1}\psi^\dagger_{\alpha\sigma}\psi^\dagger_{\beta\sigma}\psi_{\alpha\sigma}\psi_{\beta\sigma}\nonumber\\
&+g_{p2}\psi^\dagger_{\alpha\sigma}\psi^\dagger_{\beta\sigma}\psi_{\beta\sigma}\psi_{\alpha\sigma}
+g_{p3}\psi^\dagger_{\alpha\sigma}\psi^\dagger_{\alpha\sigma}\psi_{\beta\sigma}\psi_{\beta\sigma}\Big)\nonumber\\
H_{\mathrm{int},a}=\;&\frac{1}{2}\sum_{\sigma\neq\sigma'}\sum_{\alpha\neq\beta}\Big(g_{a1}\psi^\dagger_{\alpha\sigma}\psi^\dagger_{\beta\sigma'}\psi_{\alpha\sigma'}\psi_{\beta\sigma}\label{eq:EM_HP}\\
&+g_{a2}\psi^\dagger_{\alpha\sigma}\psi^\dagger_{\beta\sigma'}\psi_{\beta\sigma'}\psi_{\alpha\sigma}
+g_{a3}\psi^\dagger_{\alpha\sigma}\psi^\dagger_{\alpha\sigma'}\psi_{\beta\sigma'}\psi_{\beta\sigma}\Big)\nonumber\\
&+\frac{1}{2}\sum_{\sigma\neq\sigma'}\sum_{\alpha}g^{\alpha}_{a4}\psi^\dagger_{\alpha\sigma}\psi^\dagger_{\alpha\sigma'}\psi_{\alpha\sigma'}\psi_{\alpha\sigma}, \label{eq:EM_HA}
\end{align}
with $\psi_{\alpha\sigma}$ the fermion at $\bm K_{\alpha,\sigma}$ and the momentum sums $|\bm{q}|<\bm{q}_c$ inside each patch omitted, where $q_c$ is radius of the patch.
The $g_{p3}$ and $g_{a1}$ processes carry the residual momentum $2(\kappa_X-\kappa_Y)$. They are allowed for $t'=0$, where $\kappa_X=\kappa_Y$, and are cut off otherwise. Projecting the on-site repulsion $U$ and the nearest-neighbor repulsion $V$ onto the patches gives
\begin{align}
&g_{a2}(0)=g^X_{a4}(0)=g^Y_{a4}(0)=U+4V,\nonumber\\
&g_{a1}(0)=g_{a3}(0)=U-4V,\qquad -g_{p1}(0)=g_{p2}(0)=4V,\nonumber\\
&g_{p3}(0)=g^X_{p4}(0)=g^Y_{p4}(0)=0. \label{eq:initial_condition}
\end{align}
The equal-spin couplings $g_{p3}$ and $g^\alpha_{p4}$ vanish because their direct and exchange terms cancel.

\textit{RG equations.}---We take the $X$-patch pp bubble of Fig.~\ref{fig:1}(d) as the RG time and encode the other two bubbles in the kernels
\begin{equation}
y=-\Pi^X_{pp},\quad d(y)=-\frac{d\Pi^Y_{pp}}{dy},\quad \tilde d(y)=\frac{d\Pi^{XY}_{ph}}{dy}, \label{eq:EM_y}
\end{equation}
which we evaluate numerically. 
Both kernels equal one for $T\gg\Delta_{\mathrm{VH}}$ and $d(y)$ decreases below $\Delta_{\mathrm{VH}}$, where the neglected single-logarithmic channels become comparable to $\Pi^{Y}_{pp}$.
The flow equations of 10 couplings are given by 
\begin{subequations}
\begin{align}
\dot g_{p1}&=-\tilde d\big(g_{a1}^2+g_{a3}^2+g_{p1}^2-2g_{p1}g_{p2}-g_{p3}^2\big),\\
\dot g_{p2}&=\tilde d\big(g_{p2}^2+g_{p3}^2\big),\\
\dot g_{p3}&=-g_{p3}\big(g^X_{p4}+d\,g^Y_{p4}\big)+2\tilde d\big(2g_{p2}g_{p3}-g_{a1}g_{a3}\big),\\
\dot g^X_{p4}&=-d\,g_{p3}^2-\big(g^X_{p4}\big)^2,\\
\dot g^Y_{p4}&=-g_{p3}^2-d\big(g^Y_{p4}\big)^2,\\
\dot g_{a1}&=-2\tilde d\,g_{a1}\big(g_{p1}-g_{p2}\big),\\
\dot g_{a2}&=\tilde d\big(g_{a2}^2+g_{a3}^2\big),\\
\dot g_{a3}&=-g_{a3}\big(g^X_{a4}+d\,g^Y_{a4}\big)+2\tilde d\,g_{a3}\big(g_{a2}-g_{p1}+g_{p2}\big),\\
\dot g^X_{a4}&=-d\,g_{a3}^2-\big(g^X_{a4}\big)^2,\\
\dot g^Y_{a4}&=-g_{a3}^2-d\big(g^Y_{a4}\big)^2,
\end{align}
\label{eq:RG_flow}
\end{subequations}
with $\dot g_i\equiv dg_i/dy$. 
They can be read out from the one-loop Feynman diagrams given in Fig~\ref{fig:feynman-rules} in SM.

\textit{Test vertices and exponents.}---We add infinitesimal test vertices in the competing channels,
\begin{subequations}
\label{eq:EM_vertex}
\begin{align}
H_{\mathrm{CDW}}&=\Delta_{\mathrm{CDW}}\textstyle\sum_{s}\psi^\dagger_{X,s}\psi_{Y,s}+\mathrm{H.c.},\\
H_{\mathrm{SDW-l}}&=m_z\big(\psi^\dagger_{X,\uparrow}\psi_{Y,\uparrow}-\psi^\dagger_{X,\downarrow}\psi_{Y,\downarrow}\big)+\mathrm{H.c.},\\
H_{\mathrm{SDW-t}}&=m_+\psi^\dagger_{X,\uparrow}\psi_{Y,\downarrow}+m_-\psi^\dagger_{X,\downarrow}\psi_{Y,\uparrow}+\mathrm{H.c.},\\
H_{\mathrm{SC}}&=\textstyle\sum_{\alpha}\Delta_\alpha\psi^\dagger_{\alpha,\downarrow}\psi^\dagger_{\alpha,\uparrow}+\mathrm{H.c.}
\end{align}
\end{subequations}
The flow equations of the vertices can be derived by their one-loop corrections. 
For example, the flow equation of SC vertices are
\begin{equation}
\frac{d}{dy}\begin{pmatrix}\Delta_X\\ \Delta_Y\end{pmatrix}
=-\begin{pmatrix}g^X_{a4} & d_c g_{a3}\\ g_{a3} & d_c g^Y_{a4}\end{pmatrix}
\begin{pmatrix}\Delta_X\\ \Delta_Y\end{pmatrix}, 
\label{eq:SC_vertex_flow}
\end{equation}
which can be understood via vertex diagrams in Fig.~\ref{fig:vertex-diagrams}. Near $y_c$, we substitute $g_i\simeq G_i/(y_c-y)$ and evaluate the kernels at $d_c=d(y_c)$ and $\tilde d_c=\tilde d(y_c)$. Each vertex then obeys $d\Gamma/dy=\alpha\,\Gamma/(y_c-y)$ so that $\Gamma \sim(y_c-y)^{-\alpha}$, and its susceptibility scales as $\chi\sim(y_c-y)^{1-2\alpha}$. The SC vertices satisfy
\begin{equation}
\frac{d}{dy}\begin{pmatrix}\Delta_X\\ \Delta_Y\end{pmatrix}
=-\frac{1}{y_c-y}\begin{pmatrix}G^X_{a4} & d_cG_{a3}\\ G_{a3} & d_cG^Y_{a4}\end{pmatrix}
\begin{pmatrix}\Delta_X\\ \Delta_Y\end{pmatrix}, \label{eq:SC_vertex_traject}
\end{equation}
which gives
\begin{align}
\alpha_{\mathrm{SC},\pm}=&-\frac{1}{2}\big(G^X_{a4}+d_cG^Y_{a4}\big)\nonumber\\
&\pm\frac{1}{2}\sqrt{\big(G^X_{a4}-d_cG^Y_{a4}\big)^2+4d_cG_{a3}^2}. \label{eq:EM_aSC}
\end{align}
The fixed-trajectory equations give $G^X_{a4}=d_cG^Y_{a4}$. Then $\alpha_{\mathrm{SC},+}=-G^X_{a4}+\sqrt{d_c}\,|G_{a3}|$, and its eigenvector satisfies Eq.~\eqref{eq:gap_ratio}, $\Delta_Y/\Delta_X=-\mathrm{sgn}(G_{a3})\,d_c^{-1/2}$. The density-wave exponents are
\begin{align}
\alpha^{(r/i)}_{\mathrm{CDW}}&=\tilde d_c\big(G_{p2}-G_{p1}-G_{a1}\mp G_{a3}\big),\nonumber\\
\alpha^{(r/i)}_{\mathrm{SDW-l}}&=\tilde d_c\big(G_{p2}-G_{p1}+G_{a1}\pm G_{a3}\big),\nonumber\\
\alpha_{\mathrm{SDW-t},\pm}&=\tilde d_c\big(G_{a2}\pm G_{a3}\big), \label{eq:EM_aDW}
\end{align}
where $(r)$ and $(i)$ label the two relative-phase combinations of a vertex and its complex conjugate. All density-wave exponents carry the factor $\tilde d_c$, which is suppressed below $\Delta_{\rm VH}$, while the superconducting exponents are not. This is why superconductivity wins at finite $\lambda$. The two-sublattice model follows the same procedure with eight patches and 22 couplings and $y$ chosen according to the leading pp bubble at the considered Van Hove fillings (See SM~\cite{SM}).

\onecolumngrid
\newpage
\clearpage
\setcounter{equation}{0}
\setcounter{figure}{0}
\setcounter{table}{0}
\setcounter{page}{1}
\setcounter{section}{0}

\maketitle 
\makeatletter

\setcounter{figure}{0}
\setcounter{equation}{0}
\setcounter{table}{0}
\setcounter{section}{0}
\renewcommand{\thefigure}{S\arabic{figure}}
\renewcommand{\theequation}{S\arabic{equation}}
\renewcommand{\thetable}{S\arabic{table}}
\renewcommand{\theHfigure}{S\arabic{figure}}
\renewcommand{\theHequation}{S\arabic{equation}}
\renewcommand{\theHtable}{S\arabic{table}}
\renewcommand{\theHsection}{S\arabic{section}}

\onecolumngrid

\begin{center}
\vspace{10pt}
\textbf{\large Supplemental Material for \\``Singlet-triplet superconductivity from repulsive interactions in a $p$-wave magnet''}
\end{center} 
\begin{center}
{Hanbit Oh$^{1,\textcolor{red}{*}}$, Laura Classen$^{2,3}$, and Daniel F. Agterberg$^{1,\textcolor{red}{\dagger}}$}

\vspace{5pt}
\end{center}
\tableofcontents

\section{One-sublattice model}

\subsection{Model}
~\label{sec:1sub_model}
We consider a simple one-sublattice model in the presence of $p$-wave magnetic background on a square lattice.
The noninteracting Hamiltonian is
\begin{align}
H_{1}(\bm{k})
    &=
    \left[
        -2t(\cos k_x+\cos k_y)
        -4t'\cos k_x\cos k_y
        -\mu
    \right]\sigma_{0}
    +
    \lambda f(\bm{k})\,\sigma_{z},
    \label{eq:h0}
\end{align}
with $f(\bm{k})=\sin k_x$. Here, $t$ and $t'$ are the nearest-
and next-nearest-neighbor hopping amplitudes, respectively, $\mu$
is the chemical potential, and $\lambda$ parametrizes the odd-parity
spin splitting. 
For simplicity, we set $t'=0$ throughout paper.
Then the spin-resolved dispersions are
\begin{align}
   E_{s}(\bm{k})
    &=
    -2t(\cos k_x+\cos k_y)
    +s\lambda\sin k_x
    -\mu,
    \label{eq:lattice_dispersion}
\end{align}
where $s=\pm1$. 
Crucially, $H_1$ is invariant under the time-reversal symmetry $\mathcal{T}=i\sigma_y\mathcal{K}$ and the symmetry enforces
\begin{align}
    E_s(\bm{k})=E_{-s}(-\bm{k}).
\end{align}

\subsubsection{Patch model}
At $\lambda=0$, the spin-degenerate saddle points are located at
the two time-reversal-invariant momenta (TRIM),
\begin{align}
    X=(\pi,0),
    \qquad
    Y=(0,\pi).
\end{align}
In case of odd-parity magnet, a finite $\lambda$ shifts the saddle points originating from $X$ and $Y$ away from these high-symmetry points. 
This is because an odd-parity form factor is also odd about any TRIM point $\bm K_0$,
$f(\bm K_0+\bm q)=-f(\bm K_0-\bm q)$, since $-\bm K_0=\bm K_0\ (\mathrm{mod}\ \bm G)$.
Its gradient is therefore even about $\bm K_0$ and does not generally vanish there.
By contrast, for an even-parity form factor the gradient is odd about $\bm K_0$
and must vanish at every TRIM point.

The spin-resolved saddle-point momenta are then
\begin{align}
    \bm{K}_{X,s}
    =
    \left(
        \pi-s\bm{\kappa}_X,
        0
    \right),
    &\quad
    \bm{K}_{Y,s}
    =
    \left(
        -s\bm{\kappa}_Y,
        \pi
    \right).
    \label{eq:K_VHS}
\end{align}
with 
\begin{align}
\bm{\kappa}_X=\bm{\kappa}_Y\equiv \bm{\kappa}=\tan^{-1}\frac{\lambda}{2t}.
\label{eq:1sub_kappa}
\end{align}
Thus, components with the same spin shift in the same direction by the same amount, whereas components with opposite spins shift in opposite directions along ($k_x$) within each patch.

Moreover, the Van Hove energies are independent of spin within each patch $\alpha=X,Y$, but the two patch energies are generically different, $E_X^{\mathrm{VH}}
    \neq
    E_Y^{\mathrm{VH}}$ for $\lambda\neq0$.  
    \begin{align}
    E_X^{\mathrm{VH}}
    =
    -2t+\sqrt{4t^2+\lambda^2},
    &
    \quad
    E_Y^{\mathrm{VH}}
    =
    2t-\sqrt{4t^2+\lambda^2}.
    \label{eq:E_VH}
\end{align}
Their splitting $ \Delta\equiv 
    E_X^{\mathrm{VH}}-E_Y^{\mathrm{VH}}$ is
\begin{align}
    \Delta
    &=
    2\sqrt{4t^2+\lambda^2}-4t
    =
    \frac{\lambda^2}{2t}
    +\mathcal{O}(\lambda^4).
    \label{eq:Delta_XY}
\end{align}
Therefore, at finite $\lambda$, a single chemical potential cannot
place both saddle-point energies exactly at the Fermi level. 
In the main-text, we mainly set $\mu=E_X^{\mathrm{VH}}$.

Writing $\bm{k}=\bm{K}_{\alpha,s}+\bm{q}$,
the quadratic patch dispersions at each patch $\alpha=X,Y$
are
\begin{align}
   \epsilon_{X,s}(\bm{q})
    &=
    -\frac{\sqrt{4t^2+\lambda^2}}{2}q_x^2
    +tq_y^2
    +\mathcal{O}(q^3),
    % \label{eq:patch_dispersion_X}
    \nonumber
    \\
     \epsilon_{Y,s}(\bm{q})
    &=
    -\Delta
    +\frac{\sqrt{4t^2+\lambda^2}}{2}q_x^2
    -tq_y^2
    +\mathcal{O}(q^3).
    \label{eq:patch_dispersion}
\end{align}
This satisfies $\epsilon_{\alpha,s}(\bm{q})=\epsilon_{\alpha,s}(-\bm{q})=\epsilon_{\alpha,s}(\bm{q})$. 
The quadratic dispersions at the $X$ and $Y$ patches have exactly opposite curvatures, so that their principal axes are rotated by $\pi/2$ with respect to each other.

At $\mu=E_X^{\mathrm{VH}}$, the 
density of states at $X$ patch 
are logarithmically divergent with the prefactor $\rho_0$,
\begin{align}
D_X(E)
&=
\rho_0\log\frac{\Lambda}{|E|},
\quad
\rho_0
=
\frac{1}{
8\pi^2
\sqrt{
\frac{t}{2}\sqrt{4t^2+\lambda^2}
}
}.
\end{align}

\subsubsection{Bare susceptibilities}

Fermion loops in the RG calculation are associated with bare susceptibilities in the particle-hole and particle-particle channels.
The low-energy dispersions are independent of the spin index, $\epsilon_{\alpha,s}(\bm q)=\epsilon_\alpha(\bm q)$ [See Eq.\eqref{eq:patch_dispersion}]. Consequently, the bare susceptibilities depend only on the patch indices $\alpha,\beta=X,Y$ but independent on spin index, so we hereafter drop the spin labels $\sigma,\sigma'$.
We define
\begin{align}
\Pi_{pp}^{\alpha\beta}(i\Omega_m;T)
&\equiv
\Pi_{pp}^{\alpha,\sigma;\beta,\sigma'}
\left(
i\Omega_m,
\bm Q=\bm K_{\alpha,\sigma}+\bm K_{\beta,\sigma'}
\right)
\nonumber\\
&=
-T\sum_{\omega_n}\int_{\bm q}
G_{\alpha}(i\omega_n,\bm q)
G_{\beta}(i\Omega_m-i\omega_n,-\bm q)
\nonumber\\
&=
-\int_{\bm q}
\frac{
n_F\!\left(\epsilon_\alpha(\bm q)\right)
-
n_F\!\left(-\epsilon_\beta(\bm q)\right)
}{
i\Omega_m-\epsilon_\alpha(\bm q)-\epsilon_\beta(\bm q)
},
\label{eq:pp-bubble}
\end{align}
and
\begin{align}
\Pi_{ph}^{\alpha\beta}(i\Omega_m;T)
&\equiv
\Pi_{ph}^{\alpha,\sigma;\beta,\sigma'}
\left(
i\Omega_m,
\bm Q=\bm K_{\alpha,\sigma}-\bm K_{\beta,\sigma'}
\right)
\nonumber\\
&=
-T\sum_{\omega_n}\int_{\bm q}
G_{\alpha}(i\omega_n,\bm q)
G_{\beta}(i\omega_n+i\Omega_m,\bm q)
\nonumber\\
&=
-\int_{\bm q}
\frac{
n_F\!\left(\epsilon_\alpha(\bm q)\right)
-
n_F\!\left(\epsilon_\beta(\bm q)\right)
}{
i\Omega_m+\epsilon_\alpha(\bm q)-\epsilon_\beta(\bm q)
},
\label{eq:ph-bubble}
\end{align}
where $\int_{\bm q}\equiv\int\frac{d^2q}{(2\pi)^2}$ notation and $\epsilon_{\alpha}(\bm{q})=\epsilon_{\alpha}(-\bm{q})$ are used.
Fig.~\ref{fig:1}(d) in main-text is obtained by evaluating Eqs.~\eqref{eq:pp-bubble} and \eqref{eq:ph-bubble} at $\Omega_m=0$ as a function of $T$. 
In practice, we evaluate the bubbles numerically using the full lattice dispersion within a disk of radius $q_c=0.9/a$ around each saddle point.

Among the possible one-loop bubbles, we retain the intra-patch pp and inter-patch ph bubbles because they constitute the leading logarithmically divergent channels, following a similar treatment with Ref.~\cite{Rao2025vanhovesingularites}.
In particular, in the $\lambda\to0$ limit, they exhibit the
same $\log^2$ scaling.
If we take $\mu=E_X$, and introducing the energy scale
$T$, the susceptibilities to leading logarithmic divergence are
\begin{subequations}
\begin{align}
\Pi_{pp}^{X}(T)
&\simeq
-\frac{\rho_0}{2}\log^2\left(\frac{\Lambda}{T}\right),
\label{eq:Pi-pp-X}
\\
\Pi_{pp}^{Y}(T)
&\simeq
\begin{cases}
\displaystyle
-\frac{\rho_0}{2}\log^2\left(\frac{\Lambda}{T}\right),
& T>\Delta,
\\[2mm]
\displaystyle
-\rho_0 \log \left(\frac{\Lambda}{T}\right)
\log \left(\frac{\Lambda}{\Delta}\right),
& T<\Delta,
\end{cases}
\label{eq:Pi-pp-Y}
\\
\Pi_{ph}^{XY}(T)
&\simeq
\frac{\rho_0}{2}
\log^2\left(\frac{\Lambda}{\max\{T,\Delta\}}\right).
\label{eq:Pi-ph-XY}
\end{align}
\end{subequations}
The $X$-patch Cooper bubble retains the full double-logarithmic divergence, originating from the Cooper logarithm and the Van Hove
density-of-states logarithm. 
For the shifted $Y$-patch VHS, the latter is cut off at $\Delta$. Hence, for $T\ll\Delta$, only the Cooper logarithm continues to grow. 
The interpatch particle-hole bubble is likewise cut off at $\Delta$ and saturates below this scale.

Accordingly, in the RG analysis below, we retain the
intra-patch pp bubbles and the inter-patch ph bubbles for
both parallel- and antiparallel-spin configurations on an
equal footing. Different spin configurations generally carry
different external momenta, even though their bubble
functions are identical. Thus, in the
$(X_\uparrow,X_\downarrow,Y_\uparrow,Y_\downarrow)$
patch basis, the retained bubble functions can be represented
as
\begin{equation}
\bm\Pi_{pp}
=
\begin{pmatrix}
\Pi_{pp}^{X} & 0\\
0 & \Pi_{pp}^{Y}
\end{pmatrix}
\otimes 
\begin{pmatrix}
1&1\\
1&1
\end{pmatrix},
\qquad
\bm\Pi_{ph}
=
\begin{pmatrix}
0 & \Pi_{ph}^{XY}\\
\Pi_{ph}^{XY} & 0
\end{pmatrix}
\otimes \begin{pmatrix}
1&1\\
1&1
\end{pmatrix}.
\label{eq:susceptibility-matrices}
\end{equation}

\subsection{RG analysis}
~\label{sec:1sub_rg_analysis}
We construct a patch RG that keeps only fermions near the four spin-resolved saddle
points, which dominate the density of states. The low-energy theory then contains ten
distinct couplings, which we classify into parallel- and antiparallel-spin scattering.
The parallel-spin part is
\begin{align}
H_{\mathrm{int},p}
=&\frac{1}{2}\sum_{\sigma=\pm}\sum_{\alpha\neq\beta}
\Big(
g_{p1}\psi_{\alpha,\sigma}^{\dagger}\psi_{\beta,\sigma}^{\dagger}\psi_{\alpha,\sigma}\psi_{\beta,\sigma}
+g_{p2}\psi_{\alpha,\sigma}^{\dagger}\psi_{\beta,\sigma}^{\dagger}\psi_{\beta,\sigma}\psi_{\alpha,\sigma}
\nonumber\\
&+g_{p3}\psi_{\alpha,\sigma}^{\dagger}\psi_{\alpha,\sigma}^{\dagger}\psi_{\beta,\sigma}\psi_{\beta,\sigma}
\Big)
+\frac{1}{2}\sum_{\sigma=\pm}\sum_{\alpha=X,Y}
g_{p4}^{\alpha}\psi_{\alpha,\sigma}^{\dagger}\psi_{\alpha,\sigma}^{\dagger}\psi_{\alpha,\sigma}\psi_{\alpha,\sigma},
\label{eq:Hint-parallel-pwave}
\end{align}
and the antiparallel-spin part is
\begin{align}
H_{\mathrm{int},a}
=&\frac{1}{2}\sum_{\sigma\neq\sigma'}\sum_{\alpha\neq\beta}
\Big(
g_{a1}\,\psi_{\alpha,\sigma}^{\dagger}\psi_{\beta,\sigma'}^{\dagger}\psi_{\alpha,\sigma'}\psi_{\beta,\sigma}
+g_{a2}\,\psi_{\alpha,\sigma}^{\dagger}\psi_{\beta,\sigma'}^{\dagger}\psi_{\beta,\sigma'}\psi_{\alpha,\sigma}
\nonumber\\
&+g_{a3}\,\psi_{\alpha,\sigma}^{\dagger}\psi_{\alpha,\sigma'}^{\dagger}\psi_{\beta,\sigma'}\psi_{\beta,\sigma}
\Big)
+\frac{1}{2}\sum_{\sigma\neq\sigma'}\sum_{\alpha=X,Y}
g_{a4}^{\alpha}\,\psi_{\alpha,\sigma}^{\dagger}\psi_{\alpha,\sigma'}^{\dagger}\psi_{\alpha,\sigma'}\psi_{\alpha,\sigma}.
\label{eq:Hint-antiparallel-pwave}
\end{align}
The displacement of the Van Hove points imposes an additional momentum-conservation
constraint. 
Defining $\bm K_{\alpha,\sigma}=\bm K_\alpha+\sigma\bm\kappa_\alpha$, the $g_{p3}$ and $g_{a1}$ processes carry the residual momentum $\Delta\bm K=2(\bm\kappa_X-\bm\kappa_Y)$, i.e., the difference between the total incoming and outgoing momenta, so they
are cut off at low energies whenever $\bm\kappa_X\neq\bm\kappa_Y$. For $t'=0$, as in Eq.~\eqref{eq:1sub_kappa}, $\bm\kappa_X=\bm\kappa_Y$
and both processes must be kept. 
Since couplings vanish on the fixed trajectory throughout Fig.~\ref{fig:3}(b), whether removing them does not change the leading instability or its exponent.

\subsubsection{RG equations}
We use the $X$-patch Cooper bubble as the RG time,
\begin{equation}
y=-\Pi^X_{pp},\qquad
d(y)=\frac{-d\Pi^Y_{pp}}{dy},\qquad
\tilde d(y)=\frac{d\Pi^{XY}_{ph}}{dy},
\end{equation}
so that $y\simeq(\rho_0/2)\log^2(\Lambda/T)$ grows monotonically as $T$ is lowered.
The kernels $d$ and $\tilde d$ measure how fast the $Y$-patch Cooper channel and the
$X$--$Y$ particle--hole channel grow relative to $y$. 

\begin{figure}[tb]
\centering
%--- (d) ---
\begin{tikzpicture}[baseline]\begin{scope}[scale=1.5]
 \node[above] at(0,.9) {(a)};
 \draw[thick,densely dashed](.3,-0.45)--(.3,.45);
 \draw[thick,densely dashed](-.3,-0.45)--(-.3,.45);
 \draw[very thick,arc arrow=to pos 0.7 with length 2mm](-.9,.45)--(-.3,.45);
 \draw[very thick,arc arrow=to pos 0.7 with length 2mm](-.3,.45)--(.3,.45);
 \draw[very thick,arc arrow=to pos 0.7 with length 2mm](.3,.45)--(.9,.45);
 \draw[very thick,arc arrow=to pos 0.7 with length 2mm](-.9,-.45)--(-.3,-.45);
 \draw[very thick,arc arrow=to pos 0.7 with length 2mm](-.3,-.45)--(.3,-.45);
 \draw[very thick,arc arrow=to pos 0.7 with length 2mm](.3,-.45)--(.9,-.45);
 % \node[below] at(0,-.6) {$-\Pi_{pp}(0)$};
 %  \node at (-.9-0.1,.45) {$1$};
 %  \node at (-.9-0.1,-.45) {$2$};
 %  \node at (+.9+0.1,-.45) {$3$};
 %  \node at (+.9+0.1,.45) {$4$};
\end{scope}\end{tikzpicture}\ 
\hspace{20pt}
%--- (a) ---
\begin{tikzpicture}[baseline]\begin{scope}[scale=1.5]
 \node[above] at(0,.9) {(b)};
 \draw[thick,densely dashed](0,.25)--(0,.45);
 \draw[thick,densely dashed](0,-.25)--(0,-.45);
 \draw[very thick,arc arrow=to pos 0.6 with length 2mm] (0,0.25) arc (90:270:0.25);
 \draw[very thick,arc arrow=to pos 0.6 with length 2mm] (0,-0.25) arc (-90:90:0.25);
 \draw[very thick,arc arrow=to pos 0.6 with length 2mm](-.9,.45)--(0,.45);
 \draw[very thick,arc arrow=to pos 0.6 with length 2mm](0,.45)--(.9,.45);
 \draw[very thick,arc arrow=to pos 0.6 with length 2mm](-.9,-.45)--(0,-.45);
 \draw[very thick,arc arrow=to pos 0.6 with length 2mm](0,-.45)--(.9,-.45);
 % \node[below] at(0,-.6) {$-\Pi_{ph}(Q)$};
 %   \node at (-.9-0.1,.45) {$1$};
 %  \node at (-.9-0.1,-.45) {$2$};
 %  \node at (+.9+0.1,-.45) {$3$};
 %  \node at (+.9+0.1,.45) {$4$};
\end{scope}\end{tikzpicture}\ 
\hspace{20pt}
%--- (b) ---
\begin{tikzpicture}[baseline]\begin{scope}[scale=1.5]
 \node[above] at(0,.9) {(c)};
 \draw[thick,densely dashed](0,-0.45)--(0,.45);
 \draw[thick,dashed] (0.45,0.45) arc (0:180:0.45);
 \draw[very thick,arc arrow=to pos 0.7 with length 2mm](-.9,.45)--(-.45,.45);
 \draw[very thick,arc arrow=to pos 0.7 with length 2mm](-.45,.45)--(0,.45);
 \draw[very thick,arc arrow=to pos 0.7 with length 2mm](0,.45)--(.45,.45);
 \draw[very thick,arc arrow=to pos 0.7 with length 2mm](.45,.45)--(.9,.45);
 \draw[very thick,arc arrow=to pos 0.6 with length 2mm](-.9,-.45)--(0,-.45);
 \draw[very thick,arc arrow=to pos 0.6 with length 2mm](0,-.45)--(.9,-.45);
\end{scope}\end{tikzpicture}\ 
\hspace{20pt}
\begin{tikzpicture}[baseline]\begin{scope}[scale=1.5]
 \node[above] at(0,.9) {(d)};
 \draw[thick,densely dashed](-.3,-0.45)--(.3,.45);
 \draw[thick,densely dashed](.3,-0.45)--(-.3,.45);
 \draw[very thick,arc arrow=to pos 0.7 with length 2mm](-.9,.45)--(-.3,.45);
 \draw[very thick,arc arrow=to pos 0.7 with length 2mm](-.3,.45)--(.3,.45);
 \draw[very thick,arc arrow=to pos 0.7 with length 2mm](.3,.45)--(.9,.45);
 \draw[very thick,arc arrow=to pos 0.7 with length 2mm](-.9,-.45)--(-.3,-.45);
 \draw[very thick,arc arrow=to pos 0.7 with length 2mm](-.3,-.45)--(.3,-.45);
 \draw[very thick,arc arrow=to pos 0.7 with length 2mm](.3,-.45)--(.9,-.45);
\end{scope}\end{tikzpicture}
\vspace{10pt}
\caption{\textbf{One-loop diagrams of the parquet RG.} Solid lines are patch fermions and dashed lines are the couplings $g_i$, with incoming legs 1, 2 and outgoing legs 3, 4. (a) Particle-particle ladder, which contributes $-\Pi_{pp}(\bm 0)$. (b) Particle-hole bubble with a closed fermion loop, which carries an extra minus sign and contributes $-\Pi_{ph}(\bm Q)$. (c) Particle-hole vertex correction and (d) crossed particle-hole diagram, which each contribute $+\Pi_{ph}(\bm Q)$.
Diagram (c) is accompanied by its upside-down counterpart, which gives an identical contribution and hence an extra factor of 2.
}
\label{fig:feynman-rules}
\end{figure}
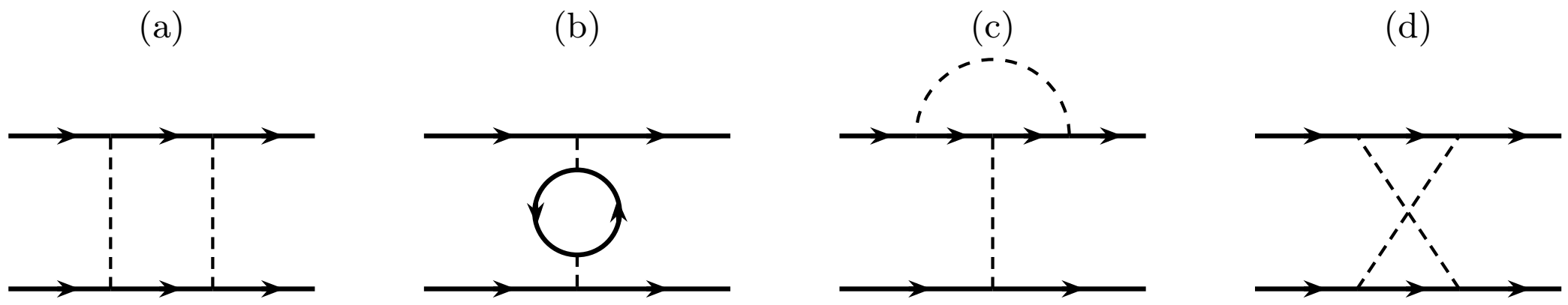

The one-loop corrections to each coupling consist of the four types of diagrams in Fig.~\ref{fig:feynman-rules}. The first ladder diagram, Fig.~\ref{fig:feynman-rules}(a) involves the Cooper bubbles, and the last three diagrams, Figs.~\ref{fig:feynman-rules}(b)--(d) involve the particle-hole bubble. Collecting all of them and differentiating with respect to $y$, we obtain
\begin{subequations}
\label{eq:S-RG}
\begin{align}
\dot g_{p1} &= -\tilde d\,\big(g_{a1}^2+g_{a3}^2+g_{p1}^2-2g_{p1}g_{p2}-g_{p3}^2\big),\\
\dot g_{p2} &= \tilde d\,\big(g_{p2}^2+g_{p3}^2\big),\\
\dot g_{p3} &= -g_{p3}\big(g^X_{p4}+d\,g^Y_{p4}\big)+2\tilde d\,\big(2g_{p2}g_{p3}-g_{a1}g_{a3}\big),\\
\dot g^X_{p4} &= -d\,g_{p3}^2-\big(g^X_{p4}\big)^2,\\
\dot g^Y_{p4} &= -g_{p3}^2-d\,\big(g^Y_{p4}\big)^2,\\
\dot g_{a1} &= -2\tilde d\,g_{a1}\big(g_{p1}-g_{p2}\big),\\
\dot g_{a2} &= \tilde d\,\big(g_{a2}^2+g_{a3}^2\big),\\
\dot g_{a3} &= -g_{a3}\big(g^X_{a4}+d\,g^Y_{a4}\big)+2\tilde d\,g_{a3}\big(g_{a2}-g_{p1}+g_{p2}\big),\\
\dot g^X_{a4} &= -d\,g_{a3}^2-\big(g^X_{a4}\big)^2,\\
\dot g^Y_{a4} &= -g_{a3}^2-d\,\big(g^Y_{a4}\big)^2,
\end{align}
\end{subequations}
where $\dot g_i\equiv dg_i/dy$.

For the initial condition, we take an on-site repulsion $U$ and a nearest-neighbor repulsion $V$,
\begin{equation}
H_{\rm int}=U\sum_{i}n_{i\uparrow}n_{i\downarrow}
+\frac{V}{2}\sum_{i,\bm\delta}\sum_{\sigma\sigma'}n_{i\sigma}n_{i+\bm\delta,\sigma'},
\end{equation}
where $\bm\delta=\pm\hat x,\pm\hat y$. In momentum space,
\begin{equation}
H_{\rm int}=\frac{1}{2N}\sum_{\bm k\bm k'\bm q}\sum_{\sigma\sigma'}
W_{\sigma\sigma'}(\bm q)\,
c^\dagger_{\bm k+\bm q,\sigma}c^\dagger_{\bm k'-\bm q,\sigma'}c_{\bm k',\sigma'}c_{\bm k,\sigma},
\qquad
W_{\sigma\sigma'}(\bm q)=U\,\delta_{\sigma,-\sigma'}+V(\bm q),
\end{equation}
with $V(\bm q)=2V(\cos q_x+\cos q_y)$.
Restricting all four momenta to the patches and comparing with
Eqs.~(\ref{eq:Hint-parallel-pwave}) and (\ref{eq:Hint-antiparallel-pwave}), each coupling
equals $W_{\sigma\sigma'}$ evaluated at its momentum transfer.
Processes $g_2$ and $g_4$ keep each electron in its patch, so $\bm q=\bm 0$.
Processes $g_1$ and $g_3$ move an electron between $\bm K_{X,\sigma}$ and $\bm K_{Y,\sigma}$,
so $\bm q=\bm K_{X,\sigma}-\bm K_{Y,\sigma}=\bm Q\equiv(\pi,\pi)$. 
With $V(\bm 0)=4V$ and $V(\bm Q)=-4V$, the antiparallel couplings are
\begin{equation}
g_{a2}(0)=g^X_{a4}(0)=g^Y_{a4}(0)
=U+V(\bm 0)=U+4V,\qquad
g_{a1}(0)=g_{a3}(0)=U+V(\bm Q)=U-4V .
\end{equation}
For parallel spins the on-site term drops out, $W_{\sigma\sigma}=V(\bm q)$, giving
\begin{equation}
g_{p2}(0)=V(\bm 0)=4V,\qquad g_{p1}(0)=V(\bm Q)=-4V .
\end{equation}
For $g_{p3}$ and $g^{\alpha}_{p4}$ the two equal-spin electrons are created (annihilated) in the
same patch. The direct and exchange terms then carry the same momentum transfer and cancel
by antisymmetry,
\begin{equation}
g_{p3}(0)=V(\bm Q)-V(\bm Q)=0,\qquad g^{\alpha}_{p4}(0)=V(\bm 0)-V(\bm 0)=0 .
\end{equation}
In summary,
\begin{equation}
\begin{aligned}
&g_{a2}(0)=g^X_{a4}(0)=g^Y_{a4}(0)=U+4V,\quad g_{a1}(0)=g_{a3}(0)=U-4V,\\
&-g_{p1}(0)=g_{p2}(0)=4V,\quad g_{p3}(0)=g^X_{p4}(0)=g^Y_{p4}(0)=0 .
\end{aligned}
\end{equation}
For the pure Hubbard model ($V=0$), only the antiparallel couplings are finite and all
equal $U$.

\subsubsection{Fixed trajectory}
\label{sec:1sub_fixed}
Near the critical scale $y_c$ all couplings diverge with fixed ratios,
\begin{equation}
g_i(y)\simeq\frac{G_i}{y_c-y},\qquad G_i\equiv\lim_{y\to y_c}(y_c-y)\,g_i(y).
\end{equation}
The set $\{G_i\}$ defines a fixed trajectory. Substituting this form into Eqs.~\eqref{eq:S-RG}, with $d\to d_c\equiv d(y_c)$ and
$\tilde d\to\tilde d_c\equiv\tilde d(y_c)$, turns the flow equations into algebraic equations
for $G_i$. 
The initial conditions enter the solution only through the critical scale $y_c$, i.e.,
through $(d_c,\tilde d_c)$, and through the sign of $g_{a3}(0)$.

\textit{Generic case, $V/U\neq1/4$.}--- We find that the flow reaches a trajectory with $G_{p2}=G_{p3}=G^{X,Y}_{p4}=G_{a1}=0$.
The condition reduces 
\begin{subequations}
\begin{align}
G_{p1}&=-\tilde d_c\big(G_{a3}^2+G_{p1}^2\big), \\
G_{a2}&=\tilde d_c\big(G_{a2}^2+G_{a3}^2\big), \\
G_{a3}&=-G_{a3}\big(G^X_{a4}+d_cG^Y_{a4}\big)+2\tilde d_c\,G_{a3}\big(G_{a2}-G_{p1}\big), \\
G^X_{a4}&=-d_cG_{a3}^2-\big(G^X_{a4}\big)^2, \\
G^Y_{a4}&=-G_{a3}^2-d_c\big(G^Y_{a4}\big)^2. \nonumber
\end{align}
\end{subequations}
The first two equations give $G_{a2}=-G_{p1}$ and the last two equations lead $G^X_{a4}=d_cG^Y_{a4}$. Finally, for $G_{a3}\neq0$ fixes
\begin{equation}
1=-2G^X_{a4}+4\tilde d_c\,G_{a2}.
\end{equation}
Therefore there are three independent relations which determine non-zero $\{G_i\}$ for given $(d_c,\tilde d_c)$.
Since $\dot g_{a3}\propto g_{a3}$, the sign of $g_{a3}$ is preserved along the flow, so
${\rm sgn}\,G_{a3}={\rm sgn}\,g_{a3}(0)={\rm sgn}(U-4V)$. The regions $V/U<1/4$ and
$V/U>1/4$ therefore lie on the same trajectory up to $G_{a3}\to-G_{a3}$.

For $d_c=\tilde d_c=1$ at $\lambda=0$ the solution is
\begin{equation}
G_{a2}=-G_{p1}=\tfrac16,\qquad G^{X}_{a4}=G^Y_{a4}=-\tfrac16,\qquad
G_{a3}=\pm\tfrac{\sqrt5}{6},
\end{equation}
with the upper (lower) sign for $V/U<1/4$ ($V/U>1/4$).
At $\lambda=0$, spin $SU(2)$ symmetry is restored, and this is the corresponding $SU(2)$-symmetric fixed trajectory.
As $\lambda$ increases, $\tilde d_c$ decreases faster than $d_c$. Then $G_{a2}$ and $-G_{p1}$
shrink, while $|G_{a3}|$ and $|G^{X}_{a4}|$ grow as in Fig.~3(b) of main-text.

\begin{figure}[tb]
    \centering
    \includegraphics[width=0.5\linewidth]{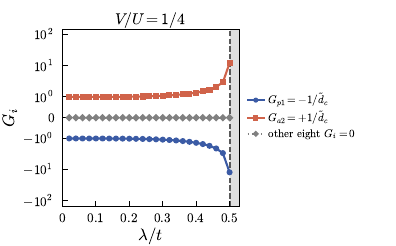}
    \hspace{-12mm}%
        \includegraphics[width=0.5\linewidth]{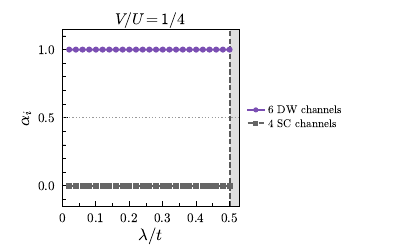}
        \vspace{-10pt}
\caption{\textbf{One-sublattice model at $V/U=1/4$.} (Left) Fixed trajectory $G_i$ and (right) scaling exponents $\alpha_i$ as functions of $\lambda$. Only $G_{p1}$ and $G_{a2}$ are nonzero, so the six density-wave channels have $\alpha=1$ and the superconducting channels have $\alpha=0$. }
\label{fig:s_1sub_G_alpha}
\end{figure}

\textit{The line $V/U=1/4$.}--- Here $g_{a1}(0)=g_{a3}(0)=U-4V=0$. Since $\dot g_{a1}\propto g_{a1}$ and $\dot g_{a3}\propto g_{a3}$,
both stay zero, and so do $g_{p3}$ and $g^{X,Y}_{p4}$. The intrapatch couplings obey
$\dot g^X_{a4}=-(g^X_{a4})^2$ and $\dot g^Y_{a4}=-d\,(g^Y_{a4})^2$ with positive initial values,
so they are marginally irrelevant. The remaining flows decouple into
\begin{equation}
\dot g_{a2}=\tilde d\,g_{a2}^2,\qquad \dot w=-\tilde d\,w^2,\qquad \dot g_{p2}=\tilde d\,g_{p2}^2,
\qquad w\equiv g_{p1}-g_{p2},
\end{equation}
with $g_{a2}(0)=2U$, $w(0)=-2U$, and $g_{p2}(0)=U$. The first two diverge together at
$\int_0^{y_c}\tilde d\,dy=1/2U$, whereas $g_{p2}$ remains finite at $y_c$. Hence, as shown in Fig.~\ref{fig:s_1sub_G_alpha}(a), we have
\begin{equation}
G_{a2}=-G_{p1}=1/\tilde d_c,\qquad \text{all other } G_i=0 .
\end{equation}
This trajectory is unstable. Linearizing Eqs.~\eqref{eq:S-RG} around it gives
$\delta\dot g_{a3}=4\,\delta g_{a3}/(y_c-y)$ and $\delta\dot g_{a1}=2\,\delta g_{a1}/(y_c-y)$.
Any deviation from $V/U=1/4$ therefore grows as $(y_c-y)^{-4}$ and $(y_c-y)^{-2}$,
faster than the couplings on the trajectory, and the flow returns to the generic trajectory.
The line $V/U=1/4$ thus separates $G_{a3}>0$ from $G_{a3}<0$.

\subsubsection{Test vertices}
To identify the physical instabilities associated with the divergent
couplings, we introduce infinitesimal test vertices in the relevant channels and follow their renormalization under the parquet RG flow.  
The leading test vertices are
\begin{subequations}
\label{eq:order_parameter_definitions}
\begin{align}
    H_{\mathrm{CDW}}
    &=
    \Delta_{\mathrm{CDW}}
    \sum_{s=\pm }
    \psi^\dagger_{X,s}\psi_{s Y,s}
    +\mathrm{H.c.},
    \label{eq:order_CDW}
    \\
    H_{\mathrm{SDW-l}}
    &=
    m_z
    \left(
\psi^\dagger_{X,\uparrow}\psi_{Y,\uparrow}
    -
\psi^\dagger_{ X,\downarrow}\psi_{ Y,\downarrow}
    \right)
    +\mathrm{H.c.},
    \label{eq:order_SDWL}
    \\
    H_{\mathrm{SDW-t}}
    &=
m_{+}\psi^\dagger_{X,\uparrow}\psi_{Y,\downarrow}
    +
m_{-}\psi^\dagger_{{X},\downarrow}\psi_{{Y},\uparrow}
    +\mathrm{H.c.},
    \label{eq:order_SDWT}
    \\
    H_{\mathrm{SC}}
    &=
    \Delta_X
\psi^\dagger_{X,\downarrow}\psi^\dagger_{X,\uparrow}
    +
\Delta_Y
\psi^\dagger_{Y,\downarrow}\psi^\dagger_{Y,\uparrow}
    +\mathrm{H.c.}.
    \label{eq:order_SC}
\end{align}
\end{subequations}
 with $\alpha=X,Y$.
 %where $\bm K_{\alpha}=\bm K_{\alpha,\uparrow}=-\bm K_{\alpha,\downarrow}$
The charge-density wave and the longitudinal spin-density wave carry
the commensurate ordering momentum $\bm Q_0=\bm{K}_{X,s}-\bm{K}_{Y,s}=(\pi,\pi)$ and therefore
break lattice translation symmetry.  They preserve the residual
spin-rotation symmetry $U(1)_{S_z}$.  The longitudinal SDW describes a
modulation of the spin density parallel to the background magnetic
axis, whereas the CDW is spin symmetric.
The transverse components $m_{+}$ and $m_{-}$ carry the incommensurate momenta $\bm Q_{+}= \bm K_{X,+}-\bm K_{Y,-}
    =
    (\pi-2\kappa,\pi),
    $ and $\bm Q_{-}  =
    (\pi+2\kappa,\pi)
    =
    -\bm Q_{+}
    \ (\mathrm{mod}\ \bm G)$, respectively.
A transverse SDW therefore breaks both lattice translation symmetry and the residual $U(1)_{S_z}$ spin-rotation symmetry.  
The superconducting vertex in has zero
center-of-mass momentum and hence does not break lattice translations,
but it spontaneously breaks the global charge $U(1)$ symmetry.

Note that four-point interaction terms in Eqs.~\eqref{eq:Hint-parallel-pwave} implicitly contain a momentum summation around each VHS within the cutoff $\Lambda$.
For example, $g_3$ or $g_4$ like couplings explicitly means 
\begin{align}
\psi^\dagger_{\alpha,\sigma}\psi^\dagger_{\alpha,\sigma'}\psi_{\beta,\sigma'}\psi_{\beta,\sigma}
\rightarrow
\sum_{\bm q,\bm k,\bm k'}
\psi^\dagger_{\alpha,\sigma}(\bm k+\bm q)\,
\psi^\dagger_{\alpha,\sigma'}(\bm k'-\bm q)\,
\psi_{\beta,\sigma'}(\bm k')\,
\psi_{\beta,\sigma}(\bm k),
\end{align}
where $\psi_{\alpha,\sigma}(\bm k)=c_\sigma(\bm K_{\alpha,\sigma}+\bm k)$.
Accordingly, the test vertex in Eqs.~\eqref{eq:order_CDW}-\eqref{eq:order_SC} also implicitly contains a momentum summation, i.e.  SC vertex implicitly means, 
\begin{equation}
H_{\rm SC}\rightarrow\sum_{|\bm k|<\Lambda}
\Delta_X\,\psi^\dagger_{X,\downarrow}(-\bm k)\,\psi^\dagger_{X,\uparrow}(\bm k)
+\Delta_Y\,\psi^\dagger_{Y,\downarrow}(-\bm k)\,\psi^\dagger_{Y,\uparrow}(\bm k)+{\rm H.c.}
\end{equation}
Therefore, $\psi^\dagger_{\bm K_{\alpha},\downarrow}\psi^\dagger_{-\bm K_{\alpha},\uparrow}$ does not show up in Eq.~\eqref{eq:order_SC}, the pairing implicitly was included in momentum sum. 
For example, in the $X$ family, this contains two opposite-spin pairs.
The first is the pair of the two VHS electrons, $P_{\downarrow\uparrow}=\psi^\dagger_{X,\downarrow}(\bm 0)\,\psi^\dagger_{X,\uparrow}(\bm 0)=c^\dagger_\downarrow(\bm K_{X,\downarrow})\,c^\dagger_\uparrow(\bm K_{X,\uparrow})$.
The second is its opposite-spin partner, $P_{\uparrow\downarrow}=\psi^\dagger_{X,\uparrow}(2\bm \kappa)\,\psi^\dagger_{X,\downarrow}(-2\bm \kappa)=c^\dagger_\uparrow(\bm K_{X,\downarrow})\,c^\dagger_\downarrow(\bm K_{X,\uparrow})$, which lies away from the VHS energy. 
Because of this energy difference, the ratio between $P_{\downarrow\uparrow}$ and $P_{\uparrow\downarrow}$ is generically not one, which implies a singlet-triplet mixture.

The one-loop RG equations for the test vertices are then
\begin{subequations}
\label{eq:test_vertex_RG}
\begin{align}
    \frac{d}{dy}
    \begin{pmatrix}
        \Delta_{\mathrm{CDW}}\\
        \Delta_{\mathrm{CDW}}^*
    \end{pmatrix}
    &=
    \widetilde d(y)
    \begin{pmatrix}
        g_{p2}-g_{p1}-g_{a1} & -g_{a3}\\
        -g_{a3} &  g_{p2}-g_{p1}-g_{a1}
    \end{pmatrix}
    \begin{pmatrix}
        \Delta_{\mathrm{CDW}}\\
        \Delta_{\mathrm{CDW}}^*
    \end{pmatrix},
    \label{eq:RG_CDW}
    \\
       \frac{d}{dy}
    \begin{pmatrix}
        m_z\\
        m_z^*
    \end{pmatrix}
    &=
    \widetilde d(y)
    \begin{pmatrix}
       g_{p2}-g_{p1}+g_{a1} & g_{a3}\\
        g_{a3} & g_{p2}-g_{p1}+g_{a1}
    \end{pmatrix}
    \begin{pmatrix}
        m_z\\
        m_z^*
    \end{pmatrix},
    \label{eq:RG_SDWL}
    \\
    \frac{d}{dy}
    \begin{pmatrix}
        m_{+}\\
        m_{-}^*
    \end{pmatrix}
    &=
    \widetilde d(y)
    \begin{pmatrix}
        g_{a2} & g_{a3}\\
        g_{a3} & g_{a2}
    \end{pmatrix}
    \begin{pmatrix}
        m_{+}\\
        m_{-}^*
    \end{pmatrix},
    \label{eq:RG_SDWT}
    \\
    \frac{d}{dy}
    \begin{pmatrix}
        \Delta_X\\
        \Delta_Y
    \end{pmatrix}
    &=
    -
    \begin{pmatrix}
        g_{a4}^{X} & d(y)g_{a3}\\
        g_{a3} & d(y)g_{a4}^{Y}
    \end{pmatrix}
    \begin{pmatrix}
        \Delta_X\\
        \Delta_Y
    \end{pmatrix}.
    \label{eq:RG_SC}
\end{align}
\end{subequations}
The complex-conjugate transverse doublet
$(m_{+}^*,m_{-})$ obeys the same equation as
Eq.~\eqref{eq:RG_SDWT}.

\begin{figure}[h]
\centering
\tikzset{source/.style={thick,decorate,decoration={zigzag,segment length=1.25mm,amplitude=0.6mm}},
         int/.style={thick,decorate,decoration={snake,segment length=1.7mm,amplitude=0.55mm}}}
%--- (a) loop in the X patch ---
\begin{tikzpicture}[baseline]\begin{scope}[scale=1.3]
 \node[above] at(-.2,1.0) {(a)};
 \draw[source](-1.45,0)--(-.85,0);
 \draw[very thick,arc arrow=to pos 0.6 with length 1.6mm](-.85,0)--(0,.75);
 \draw[very thick,arc arrow=to pos 0.6 with length 1.6mm](-.85,0)--(0,-.75);
 \draw[int](0,.75)--(0,-.75);
 \draw[very thick,arc arrow=to pos 0.6 with length 1.6mm](0,.75)--(.8,.75);
 \draw[very thick,arc arrow=to pos 0.6 with length 1.6mm](0,-.75)--(.8,-.75);
\end{scope}\end{tikzpicture}\qquad\hspace{20pt}
%--- (b) loop in the Y patch ---
\begin{tikzpicture}[baseline]\begin{scope}[scale=1.3]
 \node[above] at(-.2,1.0) {(b)};
 \draw[source](-1.45,0)--(-.85,0);
 \draw[very thick,densely dotted,arc arrow=to pos 0.6 with length 1.6mm](-.85,0)--(0,.75);
 \draw[very thick,densely dotted,arc arrow=to pos 0.6 with length 1.6mm](-.85,0)--(0,-.75);
 \draw[int](0,.75)--(0,-.75);
 \draw[very thick,arc arrow=to pos 0.6 with length 1.6mm](0,.75)--(.8,.75);
 \draw[very thick,arc arrow=to pos 0.6 with length 1.6mm](0,-.75)--(.8,-.75);
\end{scope}\end{tikzpicture}

\caption{\textbf{One-loop renormalization of the superconducting test vertex $\Delta_\alpha$.} The wavy lines are the couplings $g_i$, and solid (dotted) lines are fermions in the $X$ ($Y$) patch. The internal pair lies in the $X$ patch in (a) and in the $Y$ patch in (b), giving $\Pi^X_{pp}$ and $\Pi^Y_{pp}$, respectively.}
\label{fig:vertex-diagrams}
\end{figure}
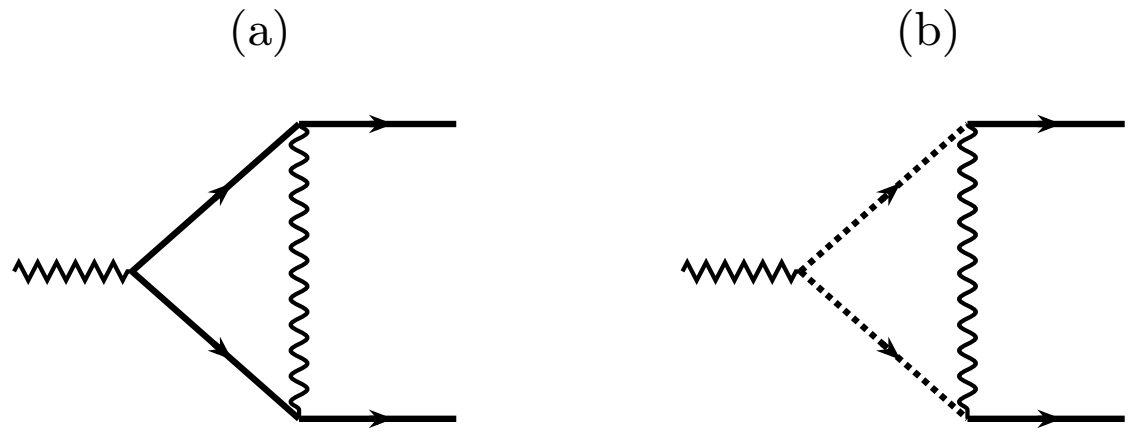

\subsubsection{Ordering instabilities}
Near the critical scale, we replace
\begin{equation}
    g_i(y)
    =
    \frac{G_i}{y_c-y},
    \qquad
    d_c=d(y_c),
    \qquad
    \widetilde d_c=\widetilde d(y_c).
    \label{eq:critical_loop_values}
\end{equation}
The vertex RG equations then take the form
\begin{equation}
    \frac{d\Gamma}{dy}
    =
    \frac{\alpha}{y_c-y}\Gamma,
    \label{eq:alpha_vertex_equation}
\end{equation}
where $\alpha$ is an eigenvalue of the corresponding critical vertex
matrix.  
Then, the susceptibility of given order parameter is given by 
\begin{align}
\chi_\Gamma \sim (y_c-y)^{1-2\alpha_\Gamma},
\end{align}
thus we need $\alpha_\Gamma > 1/2$ condition for an instability to occur.

Diagonalizing Eqs.~\eqref{eq:order_CDW} and \eqref{eq:order_SDWL} gives the real and imaginary combinations $\Phi^{(r)}=(\Phi+\Phi^*)/\sqrt{2}$ and $\Phi^{(i)}=(\Phi-\Phi^*)/(i\sqrt{2})$, respectively, where $\Phi=\Delta_{\mathrm{CDW}}$ or $m_z$. The resulting density-wave exponents are
\begin{subequations}
\label{eq:DW_exponents}
\begin{align}
    \alpha_{\mathrm{CDW}}^{(r)}
    &=
    \widetilde d_c
    \left(
    G_{p2}-G_{p1}-G_{a1}-G_{a3}
    \right),
    \label{eq:alpha_CDW_real}
    \\
    \alpha_{\mathrm{CDW}}^{(i)}
    &=
    \widetilde d_c
    \left(
    G_{p2}-G_{p1}-G_{a1}+G_{a3}
    \right),
    \label{eq:alpha_CDW_imag}
    \\
    \alpha_{\mathrm{SDW-l}}^{(r)}
    &=
    \widetilde d_c
    \left(
    G_{p2}-G_{p1}+G_{a1}+G_{a3}
    \right),
    \label{eq:alpha_SDWL_real}
    \\
    \alpha_{\mathrm{SDW-l}}^{(i)}
    &=
    \widetilde d_c
    \left(
    G_{p2}-G_{p1}+G_{a1}-G_{a3}
    \right).
    \label{eq:alpha_SDWL_imag}
    \\
    \alpha_{\mathrm{SDW-t},+}
    &=
    \widetilde d_c
    \left(
    G_{a2}+G_{a3}
    \right),
    \label{eq:alpha_SDWT_plus}
    \\
    \alpha_{\mathrm{SDW-t},-}
    &=
    \widetilde d_c
    \left(
    G_{a2}-G_{a3}
    \right),
    \label{eq:alpha_SDWT_minus}
\end{align}
\end{subequations}
The superconducting exponents are given by
\begin{equation}
    \alpha_{\mathrm{SC},\pm}
    =
    -\frac{G_{a4}^{X}+d_cG_{a4}^{Y}}{2}
    \pm
    \frac{1}{2}
    \sqrt{
        \left(
        G_{a4}^{X}-d_cG_{a4}^{Y}
        \right)^2
        +
        4d_cG_{a3}^2
    }.
    \label{eq:alpha_SC_general}
\end{equation}

In Fig.~\ref{fig:3}(a,c), we illustrate the $\lambda$ and $V/U$ dependence of $\alpha$ of vertices demonstrating that SC$_+$ state is always the leading instability, except the fine-tuned $V/U=1/4$ line. 
We illustrate the results for $V/U=1/4$ in Fig.~\ref{fig:s_1sub_G_alpha}(b).
Here $\alpha_{\mathrm{SC},+}$ denotes the larger eigenvalue. 
On the general fixed trajectory, $G^X_{a4}=d_cG^Y_{a4}$, and its eigenmode satisfies
\begin{equation}
    \Delta_Y/\Delta_X=-\,\mathrm{sgn}(G_{a3})\,d_c^{-1/2}.
\end{equation}
The relative sign of the gap on the two families is therefore set by $G_{a3}$ decided by $U-4V$. For $V/U<1/4$ ($G_{a3}>0$) the gap changes sign between $X$ and $Y$ ($d$-like, $\Delta_X\Delta_Y<0$), whereas for $V/U>1/4$ ($G_{a3}<0$) it keeps its sign (Extended $s$-like,
$\Delta_X\Delta_Y>0$).

\subsection{Results with single-log channels for $T_c<\Delta_{\rm VH}$}
~\label{sec:1sub_singlelog}

In this section, we present pRG results that include all kernels neglected in Eq.~\eqref{eq:susceptibility-matrices} within the Van Hove patch model, namely the interpatch pp susceptibility $\Pi_{\mathrm{pp}}^{XY}(\bm Q)$ and the intrapatch ph susceptibility $\Pi_{\mathrm{ph}}^\alpha(0)$ with $\alpha=X,Y$. Note that Fermi-surface regions away from the Van Hove points also can generate conventional single-logarithmic pp bubbles, but these contributions are not included in the present patch theory.
For $T>\Delta_{\rm VH}$, all three bubbles grow only as a single logarithm, $\Pi^{XY}_{pp}(\bm Q)\simeq\Pi^{\alpha\alpha}_{ph}(\bm 0)\simeq\rho_0\log(\Lambda/T)$, in contrast to the $\log^2$ growth of $\Pi^{\alpha}_{pp}(\bm 0)$ and $\Pi^{XY}_{ph}(\bm Q)$.
Therefore, for $T_c>\Delta_{\rm VH}$ it was sufficient to keep only the leading-log channels in the pRG analysis.
For $T<\Delta_{\rm VH}$, $\Pi^{Y}_{ph}(\bm 0)$ and $\Pi^{XY}_{pp}(\bm Q)$ saturate at $\simeq\rho_0\log(\Lambda/\Delta_{\rm VH})$, whereas $\Pi^{X}_{ph}(\bm 0)$ continues to grow as $\rho_0\log(\Lambda/T)$.
In this regime, $\Pi^{Y}_{pp}(\bm 0)\simeq\rho_0\log(\Lambda/\Delta_{\rm VH})\log(\Lambda/T)$ is also only singly logarithmic in $T$, and $\Pi^{XX}_{ph}(\bm 0)$ is smaller than it only by the factor $\log(\Lambda/\Delta_{\rm VH})$.
A consistent treatment for $T_c<\Delta_{\rm VH}$ should therefore include $\Pi^{XX}_{ph}(\bm 0)$ on the same footing as the leading-log channels.
For completeness, we retain all three bubbles in the flow equations below.

With the additional kernels, we additionally define 
\begin{equation}
d_0^{\alpha}(y)=\frac{d\Pi^{\alpha\alpha}_{ph}(\bm 0)}{dy},\qquad
d_2(y)=-\frac{d\Pi^{XY}_{pp}(\bm Q)}{dy},
\end{equation}
and the flow equations \eqref{eq:S-RG} are modified by the additional terms $d_{0}^X$,$d_{0}^Y$ and $d_2$: 
\begin{subequations}
\begin{align}
\dot g_{p1}&\rightarrow \dot g_{p1}-2d_2\,g_{p1}g_{p2},\\
\dot g_{p2}&\rightarrow \dot g_{p2}-d_2\big(g_{p1}^2+g_{p2}^2\big)-\big(d_0^X g^X_{a4}+d_0^Y g^Y_{a4}\big)g_{a2},\\
\dot g_{a1}&\rightarrow \dot g_{a1}-2d_2\,g_{a1}g_{a2}+\big(d_0^X g^X_{a4}+d_0^Y g^Y_{a4}\big)g_{a1},\\
\dot g_{a2}&\rightarrow \dot g_{a2}-d_2\big(g_{a1}^2+g_{a2}^2\big)+\big(d_0^X g^X_{a4}+d_0^Y g^Y_{a4}\big)\big(g_{p1}-g_{p2}\big),\\
\dot g^X_{a4}&\rightarrow \dot g^X_{a4}+d_0^X\big(g^X_{a4}\big)^2+d_0^Y\big[g_{a1}^2+2\big(g_{p1}-g_{p2}\big)g_{a2}\big],\\
\dot g^Y_{a4}&\rightarrow \dot g^Y_{a4}+d_0^Y\big(g^Y_{a4}\big)^2+d_0^X\big[g_{a1}^2+2\big(g_{p1}-g_{p2}\big)g_{a2}\big],
\end{align}
\label{eq:S-RG_singlelog} 
\end{subequations}
while $\dot g_{a3}$ is unchanged.

\begin{figure}[b]
    \centering
\includegraphics[width=0.35\linewidth]{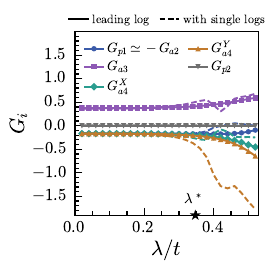}
    \hspace{10mm}
\includegraphics[width=0.35\linewidth]{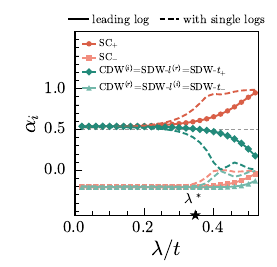}
\caption{\textbf{Effect of the single-log channels on the fixed trajectory for $U=0.1t$ and $V=0$.}
(a) Fixed-trajectory coefficients $G_i$ and (b) exponents $\alpha_i$ as functions of $\lambda$.
Solid lines with symbols are the leading-log results of Fig.~\ref{fig:3}(b,c).
Dashed lines include the single-log kernels $d_0^X$, $d_0^Y$, and $d_2$ of Eq.~\eqref{eq:S-RG_singlelog}.
In (b), the star on the horizontal axis marks $\lambda^*\simeq0.35t$, where $T_c=\Delta_{\rm VH}$.
In all cases, the leading instability is SC$_+$.
}
\label{fig:S_singlelog}
\end{figure}

The current test-vertex equations are unchanged, because the additional kernels do not enter the superconducting and density-wave vertices.
Furthermore, we do not evaluate additional test vertices in the $\bm q=0$ particle-hole and $\bm q=\bm K_X-\bm K_Y$ particle-particle channels, since they are subleading. Our analysis for $T<\Delta_{\mathrm{VH}}$ is therefore restricted to testing the robustness of the leading channels identified above the crossover scale.
We solve the modified flow equations for $U=0.1t$ and $V=0$, obtain the coefficients $G_i$ from the same fixed-trajectory analysis as above, and evaluate the exponents $\alpha_i$ from Eqs.~\eqref{eq:DW_exponents} and \eqref{eq:alpha_SC_general} with $d_c$ and $\tilde d_c$ taken at the new $y_c$.
The results are shown as dashed lines in Fig.~\ref{fig:S_singlelog}.

For small $\lambda$, the coefficients and exponents stay close to their leading-log values.
The single logarithms generate a small $G_{p2}$, and $G_{p1}=-G_{a2}$ holds only approximately.
As $\lambda$ approaches $\lambda^*$, the two sets of results begin to deviate, and beyond $\lambda^*$ the deviation becomes large, most notably for $G^Y_{a4}$.
In this regime $T_c<\Delta_{\rm VH}$, and the single-logarithmic terms are no longer parametrically small compared with the leading-log ones.
Nevertheless, $\mathrm{SC}_+$ remains the leading instability for all $\lambda$.
Its exponent is even enhanced, while the density-wave exponents fall below $1/2$.
The leading-log conclusion is therefore robust against the inclusion of the single-logarithmic channels.

\section{Two-sublattice model}
\subsection{Model}
~\label{sec:2sub_model}
We consider a $4\times4$ Hamiltonian where $\tau$ and $\sigma$ denotes the sublattice ($A,B$) and spin ($\sigma=\uparrow, \downarrow$),
\begin{align}
    H_{2}(\bm{k})
    &=
    \epsilon_0 (\bm k) \tau_0 \sigma_0
    +\epsilon_\perp (\bm k )
    \tau_x \sigma_0
    +
    \lambda f(\bm{k})\,\tau_0\sigma_{z},
    \label{eq:h0}
\end{align}
with $\epsilon_0 (\bm k) =
     -2t (\cos k _x + \cos k_y)
     -\mu$,
$\epsilon_\perp (\bm k) 
=2t_\perp \cos (k _x/2) \cos (k_y/2), 
$ and $f(\bm{k})=\sin k_x$. 
 The energy dispersion is given by $E_{\eta,\sigma}(\bm k)
=
   \epsilon_0 \pm \epsilon_x (\bm k )
    +
   \sigma \lambda f(\bm{k})$
where $\eta = \pm $ denotes the bonding/anti-bonding.

The microscopic $p$-wave magnet model breaks time reversal $\mathcal T$ but preserves $\mathcal T\tau$, with $\tau$ the half-lattice translation, and $\mathcal T C_2^\perp$, with $C_2^\perp$ a $\pi$ spin rotation about the axis perpendicular to the moments.
The effective model $H_2$, like $H_1$, is instead invariant under $\mathcal T=i\sigma_y\mathcal K$.
All three operations enforce
\begin{equation}
E_{\eta,s}(\bm k)=E_{\eta,-s}(-\bm k).
\label{eq:energy_constraint}
\end{equation}

\subsubsection{Patch model}
We now turn to analyze the VHS points. The model has 8 VHS points. The distribution of the VHSs can be understood in two steps. At $t_\perp=\lambda=0$, there are two inequivalent saddle-point momenta ($X,Y$), each has fourfold spin and sublattice degeneracy. 
Turning on $\lambda$ lifts the spin degeneracy and shifts the saddle points in momentum space, yielding four distinct VHS momenta, each still twofold degenerate in the sublattice sector. A finite $t_\perp$ further lifts the sublattice degeneracy by forming bonding and antibonding bands labeled by $\eta=\pm1$, resulting in eight distinct VHS momenta. 

To second order in $\lambda/t$ and $t_\perp/t$, the spin-up saddle points and their energies are
\begin{align}
\bm K_{X_1}&\equiv \bm K_{X_1,\uparrow}\simeq\Big(\pi-\frac{\lambda+t_\perp}{2t},\,0\Big), \qquad
E_{X_1}\simeq\frac{(\lambda+t_\perp)^2}{4t},\nonumber\\
\bm K_{X_2}&\equiv \bm K_{X_2,\uparrow}
\simeq\Big(\pi-\frac{\lambda-t_\perp}{2t},\,0\Big), \qquad
E_{X_2}\simeq\frac{(\lambda-t_\perp)^2}{4t},\nonumber\\
\bm K_{Y_{1,2}}&\equiv \bm K_{Y_{1,2},\uparrow}
\simeq\Big(-\frac{\lambda}{2t},\,\pi\mp\frac{t_\perp}{2t}\Big), \qquad
E_{Y}\simeq\frac{t_\perp^2-\lambda^2}{4t},
\label{eq:saddle_two}
\end{align}
where $X_1,Y_1$ ($X_2,Y_2$) belong to the $\eta=+$ ($\eta=-$) band and energies are measured
from $\mu=0$. The spin-down partners follow from Eq.~(\ref{eq:energy_constraint}),
$\bm K_{\alpha,\downarrow}=-\bm K_{\alpha,\uparrow}$.
Writing $\epsilon_\alpha(\bm q)= E(\bm K_\alpha+\bm q)$, the quadratic dispersions are
\begin{align}
\epsilon_{X_i}(\bm q)&\simeq E_{X_i}
-
a_{X_i}q_x^2+b_{X_i}q_y^2,\nonumber\\
\epsilon_{Y_{1}}(\bm q)&\simeq 
E_{Y}
+
a_Y q_x^2-b_Y q_y^2- c_Y\,q_xq_y,\nonumber\\
\epsilon_{Y_{2}}(\bm q)&\simeq 
E_{Y}
+
a_Y q_x^2-b_Y q_y^2+ c_Y\,q_xq_y,\nonumber\\
\end{align}
with
\begin{align}
a_{X_i}&=t-\frac{(\lambda+\eta_it_\perp)(\eta_it_\perp-2\lambda)}{16t}, \qquad
b_{X_i}=t-\frac{\eta_it_\perp(\lambda+\eta_it_\perp)}{16t},\nonumber\\
a_Y&=t+\frac{\lambda^2}{8t}-\frac{t_\perp^2}{16t}, \qquad
b_Y=t-\frac{t_\perp^2}{16t}, \qquad c_Y=\frac{\lambda t_\perp}{8t},
\end{align}
and $\eta_{1,2}=\pm1$.
The $X$ saddle points lie on the mirror line $k_y=0$, so their principal axes are fixed
along $\hat x$ and $\hat y$. The $Y_1$ and $Y_2$ saddle points are mirror images under $M_y$. Because their cross terms have opposite signs, their principal axes are rotated by $-\theta$ and $+\theta$, respectively, with $\theta\simeq\lambda t_\perp/32t^2$.

\begin{figure}
    \centering
    \includegraphics[width=0.58\linewidth]{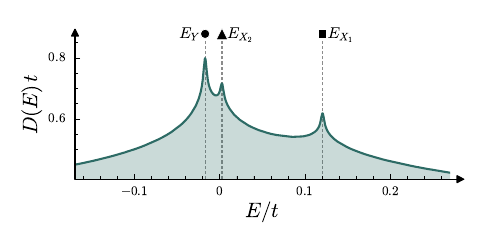}\\
      \includegraphics[width=0.3\linewidth]{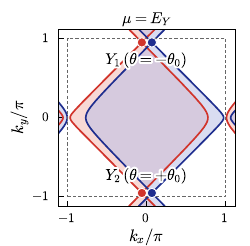}
         \hspace{-5mm}%
    \includegraphics[width=0.3\linewidth]{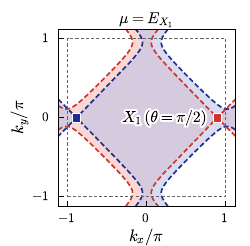}
       \hspace{-5mm}%
    \includegraphics[width=0.3\linewidth]{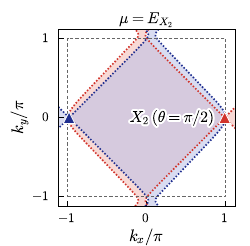}
    \vspace{-10pt}
\caption{\textbf{Van Hove singularities of the two-sublattice model.} (Top) Density of states for $\lambda/t=0.4$ and $t_\perp/t=0.3$, showing three VHS energies $E_Y$, $E_{X_2}$, and $E_{X_1}$. (Bottom) Spin-resolved constant-energy contours at each VHS energy. Red and blue denote spin up and down, and the shaded regions are occupied. Symbols mark the saddle points with $X_i,Y_i \equiv X_{i,\uparrow},Y_{i,\uparrow}=-X_{i,\downarrow},-Y_{i,\downarrow}$ and 
$\theta$ is the orientation of their principal axes, with $\theta_0\simeq\lambda t_\perp/32t^2\approx1.2\times10^{-3}\pi$. }
\label{fig:s_2sub_FS}
\end{figure}

\subsubsection{Bare susceptibility}
\begin{figure}[b]
    \centering
    \includegraphics[width=0.45\linewidth]{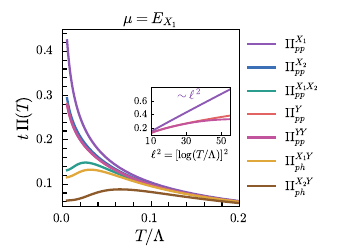}
    \hspace{-5mm}%
    \includegraphics[width=0.45\linewidth]{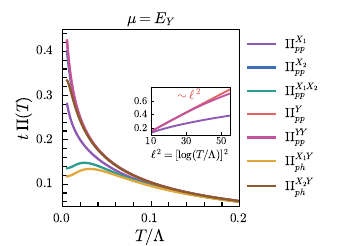}
    \vspace{-5pt}
\caption{\textbf{Bare susceptibilities of the two-sublattice model.}
The seven independent bubbles of Eq.~(\ref{eq:2sub_bubbles}) as functions of temperature for
$\lambda/t=0.4$ and $t_\perp/t=0.3$, with the chemical potential at $E_{X_1}$ (left)
and $E_{Y}$ (right).
Only the Cooper bubble of the patch at the Fermi level, $\Pi^{X_1}_{pp}$ (left) or $\Pi^{Y}_{pp}$
(right), grows as $\ell^2=[\log(T/\Lambda)]^2$. 
The off-resonant Cooper bubbles grow only as $\ell$, while
$\Pi^{X_1X_2}_{pp}$ and the particle--hole bubbles $\Pi^{X_iY}_{ph}$ saturate at low $T$.
In the right inset ($\mu=E_Y$ case), $\Pi^{YY}_{pp}$ almost follows $\Pi^{Y}_{pp}$ and deviates from it only below
$T\approx3\times10^{-3}t$ (above $\ell^2 \approx 32$), where the difference exceeds $1\%$. 
}
\label{fig:s_2sub_bubble}
\end{figure}

The bubbles $\Pi^{a\sigma;b\sigma'}_{pp}$ and $\Pi^{a\sigma;b\sigma'}_{ph}$ are defined as in
Eqs.~\eqref{eq:pp-bubble} and \eqref{eq:ph-bubble}, now for patches $a,b\in\{X_1,X_2,Y_1,Y_2\}$.
As in the one-sublattice model, we retain only the Cooper bubbles within each family and
the particle--hole bubbles between the $X_i$ and $Y$ families. 
Since $\mathcal T$ maps $\epsilon_{a\uparrow}(\bm q)$ onto $\epsilon_{a\downarrow}(-\bm q)$, and the quadratic dispersions are even in $\bm q$, the retained bubbles reduce to seven independent functions,
\begin{align}
&\Pi^{X_i}_{pp}=\Pi^{X_i\sigma;X_i\sigma'}_{pp},\qquad
\Pi^{Y}_{pp}=\Pi^{Y_j\sigma;Y_j\sigma'}_{pp},\qquad
\Pi^{X_1X_2}_{pp}=\Pi^{X_1\sigma;X_2\sigma'}_{pp},\nonumber\\
&\Pi^{YY}_{pp}=\Pi^{Y_1\sigma;Y_2\sigma'}_{pp},\qquad
\Pi^{X_iY}_{ph}=\Pi^{X_i\sigma;Y_j\sigma'}_{ph},
\label{eq:2sub_bubbles}
\end{align}
for all $\sigma,\sigma'$ and $i,j=1,2$. 
The bubbles $\Pi^{Y}_{pp}$ and $\Pi^{YY}_{pp}$ differ because the dispersions near $Y_1$ and $Y_2$ are not identical. As previously described, their principal axes are rotated by $\mp\theta$, so $\epsilon_{Y_2}(-\bm q)\neq\epsilon_{Y_1}(\bm q)$ and the $Y_1$--$Y_2$ Cooper channel loses its $\log^2$ divergence at very low energy scale. The crossover scale is controlled by the coefficient $c_Y=\lambda t_\perp/(8t)$, and we found $T\simeq3\times10^{-3}t$ for the parameters used in Fig.~\ref{fig:s_2sub_bubble}. In Fig.~\ref{fig:s_2sub_bubble}, we plot the temperature dependence of seven bubbles at two distinct cases, $\mu=E_{X_1}$ and $\mu=E_Y$.

\subsection{RG Analysis}
~\label{sec:2sub_rg_analysis}
Near $\mu \simeq E_{X_1}, E_{X_2}, E_{Y}$, we again project the interaction onto the states around the eight patch points $\bm K_{\alpha\sigma}$ with $\alpha \in \{X_1, X_2, Y_1, Y_2\}$ and $\sigma \in \{\uparrow,\downarrow\}$. Time-reversal symmetry fixes $\bm K_{\alpha\downarrow} = -\bm K_{\alpha\uparrow}$. Adopting the same channel notation as in the one-sublattice model, the possible scatterings are
\begin{align}
    g_{p,i}^{\alpha\beta},\; g_{a,i}^{\alpha\beta},\; g_{p,4}^{\alpha},\; g_{a,4}^{\alpha},
    \qquad i\in\{1,2,3\},\ \ \alpha\neq\beta \in \{X_1,X_2,Y_1,Y_2\}.
\end{align}

Without further constraints, eight patches allow $44=\big(\binom{4}{2}\times 3 + 4\big)\times 2$ couplings. Momentum conservation removes many of them. For $\alpha \neq \beta$, the processes $g_{a,1}^{\alpha\beta}$ and $g_{p,3}^{\alpha\beta}$ transfer the momentum $2(\bm K_{\alpha\uparrow} -\bm K_{\beta\uparrow})$. This is not a reciprocal lattice vector, thus these two processes are forbidden. This differs from the one-sublattice model, where $\bm \kappa_X =\bm \kappa_Y$ allows $g_{a,1}$ and $g_{p,3}$. Each pair of distinct patches thus keeps the four couplings $\{g_{p,1}^{\alpha\beta}, g_{p,2}^{\alpha\beta}, g_{a,2}^{\alpha\beta}, g_{a,3}^{\alpha\beta}\}$, and each single patch keeps $\{g_{p,4}^{\alpha}, g_{a,4}^{\alpha}\}$.

The mirror $M_y$ maps $Y_1 \leftrightarrow Y_2$ and leaves $X_1$ and $X_2$ invariant. It imposes $g^{X_i Y_1} = g^{X_i Y_2} \equiv g^{X_i Y}$ and $g_4^{Y_1} = g_4^{Y_2} \equiv g_4^{Y}$. The interpatch couplings then fall into four classes, $YY \equiv Y_1Y_2$, $X_1X_2$, $X_1Y$, and $X_2Y$, which give $16$ couplings. Together with the six intrapatch couplings $g_{p,4}^{a}$ and $g_{a,4}^{a}$ for $a = X_1, X_2, Y$, the low-energy theory contains $22$ independent couplings,
\begin{align}
H_{\text{int}} =\;& \frac{1}{2}\sum_{\alpha\neq\beta}\sum_{\sigma}
\Big[ g_{p,1}^{\alpha\beta}\, \psi^\dagger_{\alpha\sigma}\psi^\dagger_{\beta\sigma}\psi_{\alpha\sigma}\psi_{\beta\sigma}
    + g_{p,2}^{\alpha\beta}\, \psi^\dagger_{\alpha\sigma}\psi^\dagger_{\beta\sigma}\psi_{\beta\sigma}\psi_{\alpha\sigma} \Big] \nonumber\\
&+ \frac{1}{2}\sum_{\alpha\neq\beta}\sum_{\sigma\neq\sigma'}
\Big[ g_{a,2}^{\alpha\beta}\, \psi^\dagger_{\alpha\sigma}\psi^\dagger_{\beta\sigma'}\psi_{\beta\sigma'}\psi_{\alpha\sigma}
    + g_{a,3}^{\alpha\beta}\, \psi^\dagger_{\alpha\sigma}\psi^\dagger_{\alpha\sigma'}\psi_{\beta\sigma'}\psi_{\beta\sigma} \Big] \nonumber\\
&+ \frac{1}{2}\sum_{\alpha}
\Big[ \sum_{\sigma} g_{p,4}^{\alpha}\, \psi^\dagger_{\alpha\sigma}\psi^\dagger_{\alpha\sigma}\psi_{\alpha\sigma}\psi_{\alpha\sigma}
    + \sum_{\sigma\neq\sigma'} g_{a,4}^{\alpha}\, \psi^\dagger_{\alpha\sigma}\psi^\dagger_{\alpha\sigma'}\psi_{\alpha\sigma'}\psi_{\alpha\sigma} \Big].
\label{eq:Hint_2sub}
\end{align}

For the initial condition, we take an on-site repulsion $U$, an intrasublattice nearest-neighbor repulsion $V$, and an intersublattice nearest-neighbor repulsion $V_\perp$,
\begin{align}
H_{\text{int}} = U\sum_{i,l} n_{il\uparrow}n_{il\downarrow}
+ \frac{V}{2}\sum_{i,l,\bm{\delta}}\sum_{\sigma\sigma'} n_{il\sigma}n_{i+\bm{\delta},l\sigma'}
+ V_\perp\sum_{i,\bm{\delta}_\perp}\sum_{\sigma\sigma'} n_{iA\sigma}n_{i+\bm{\delta}_\perp,B\sigma'},
\end{align}
where $l=A,B$ is the sublattice index, $\bm{\delta}=\pm\hat{x},\pm\hat{y}$, and $\bm{\delta}_\perp=(\pm\hat{x}\pm\hat{y})/2$. In momentum space they give $V(\bm{q})=2V(\cos q_x+\cos q_y)$ and $V_\perp(\bm{q})=4V_\perp\cos(q_x/2)\cos(q_y/2)$. We project onto the band eigenstates $(1,\eta)/\sqrt{2}$ in sublattice space and place the patches near from $X,Y$ points to leading order in $\lambda$ and $t_\perp$. The sublattice form factor gives an overall factor $1/2$, and $V_\perp$ enters with the relative sign $\eta_\alpha\eta_\beta$ of the two bands connected by the momentum transfer. 
\begin{align}
&g_{a,2}^{\alpha\beta}(0)=g_{a,4}^{a}(0)=\tfrac{U}{2}+2V+2V_\perp,\qquad
g_{p,2}^{\alpha\beta}(0)=2V+2V_\perp,\nonumber\\
&g_{a,3}^{\alpha\beta}(0)=\tfrac{U}{2}+2V-2V_\perp,\qquad
g_{p,1}^{\alpha\beta}(0)=2V-2V_\perp,\qquad \alpha\beta\in\{X_1X_2,\,YY\},\nonumber\\
&g_{a,3}^{X_iY}(0)=\tfrac{U}{2}-2V,\qquad
g_{p,1}^{X_iY}(0)=-2V,\qquad
g_{p,4}^{a}(0)=0.
\label{eq:IC_2sub}
\end{align}
For $V=V_\perp=0$, only the antiparallel couplings are finite and all equal $U/2$.

\subsubsection{RG equations}

We take the Cooper bubble of the patch at the Fermi level as the RG time, $y=-\Pi_{pp}^{X_1}$ for $\mu=E_{X_1}$, 
$y=-\Pi_{pp}^{X_2}$ for $\mu=E_{X_2}$,
and $y=-\Pi_{pp}^{Y}$ for $\mu=E_Y$, and define the kernels
\begin{align}
d_\alpha=-\frac{d\Pi^\alpha_{pp}}{dy},\qquad
d_{YY}=-\frac{d\Pi^{YY}_{pp}}{dy},\qquad
d_{12}=-\frac{d\Pi^{X_1X_2}_{pp}}{dy},\qquad
\tilde d_i=\frac{d\Pi^{X_iY}_{ph}}{dy},
\end{align}
with $\alpha=X_1,X_2,Y$. The kernel of the resonant patch should be one. 
For example, $d_{X_1}=1$ for $\mu=E_{X_1}$, while $d_{Y}=1$ for $\mu=E_{Y}$.
The other kernels either decay slowly or even vanish at low $T$.
Fig.~\ref{fig:s_2sub_bubble} shows that $d_{YY}$ and $d_Y$ are nearly identical down to $T^\ast/t\sim10^{-3}$. Above this scale, the cross term $\mp c\,q_xq_y$ is negligible compared with $T$, and the $Y_1$--$Y_2$ Cooper bubble also grows as $\log^2$. Since $T^\ast$ is small, we set $d_{YY}=d_Y$ throughout. In later, we also find that the RG flow diverges at temperatures above $T^\ast$, so this approximation is safe.

\begin{figure}
    \centering
\includegraphics[width=0.32\linewidth]{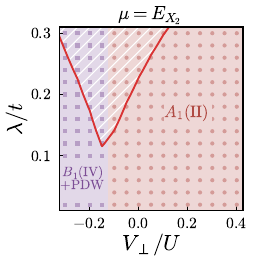}
   \vspace{-5pt}
    \caption{Leading instability of the two-sublattice model for $\mu=E_{X_2}$, with the same parameters as in Fig.~\ref{fig:4}(c,d). 
    The red line marks $\lambda^*$ satisfying $T_c=E_{X_1}-E_Y$, and the hatched region has $T_c<E_{X_1}-E_Y$. 
    The phase diagram is almost the same with $\mu=E_{Y}$ of Fig.~\ref{fig:4}(d). }
    \label{fig:S_muEX2}
\end{figure}

Collecting all one-loop diagrams, we obtain the flow equations for the 22 couplings,
\begin{subequations}
\begin{align}
\dot g_{p,1}^{YY} &= -2d_{YY}\,g_{p,1}^{YY}g_{p,2}^{YY}, \\
\dot g_{p,2}^{YY} &= -d_{YY}\big[(g_{p,1}^{YY})^2+(g_{p,2}^{YY})^2\big], \\
\dot g_{a,2}^{YY} &= -d_{YY}\,(g_{a,2}^{YY})^2, \\
\dot g_{a,3}^{YY} &= -2d_Y\,g_{a,3}^{YY}g_{a,4}^{Y} - d_{X_1}(g_{a,3}^{X_1Y})^2 - d_{X_2}(g_{a,3}^{X_2Y})^2, \\[4pt]
\dot g_{p,1}^{X_1X_2} &= -2d_{12}\,g_{p,1}^{X_1X_2}g_{p,2}^{X_1X_2}, \\
\dot g_{p,2}^{X_1X_2} &= -d_{12}\big[(g_{p,1}^{X_1X_2})^2+(g_{p,2}^{X_1X_2})^2\big], \\
\dot g_{a,2}^{X_1X_2} &= -d_{12}\,(g_{a,2}^{X_1X_2})^2, \\
\dot g_{a,3}^{X_1X_2} &= -g_{a,3}^{X_1X_2}\big(d_{X_1}g_{a,4}^{X_1}+d_{X_2}g_{a,4}^{X_2}\big) - 2d_Y\,g_{a,3}^{X_1Y}g_{a,3}^{X_2Y}, \\[4pt]
\dot g_{p,1}^{X_iY} &= -\tilde d_i\big[(g_{a,3}^{X_iY})^2+(g_{p,1}^{X_iY})^2-2g_{p,1}^{X_iY}g_{p,2}^{X_iY}\big], \\
\dot g_{p,2}^{X_iY} &= \tilde d_i\,(g_{p,2}^{X_iY})^2, \\
\dot g_{a,2}^{X_iY} &= \tilde d_i\big[(g_{a,2}^{X_iY})^2+(g_{a,3}^{X_iY})^2\big], \\
\dot g_{a,3}^{X_iY} &= -g_{a,3}^{X_iY}\big[d_{X_i}g_{a,4}^{X_i}+d_Y\big(g_{a,4}^{Y}+g_{a,3}^{YY}\big)\big]
 - d_{X_{\bar{i}}}\,g_{a,3}^{X_1X_2}g_{a,3}^{X_{\bar{i}}Y} \nonumber\\
&\quad + 2\tilde d_i\,g_{a,3}^{X_iY}\big(g_{a,2}^{X_iY}-g_{p,1}^{X_iY}+g_{p,2}^{X_iY}\big), \\[4pt]
\dot g_{a,4}^{X_i} &= -d_{X_i}(g_{a,4}^{X_i})^2 - d_{X_{\bar{i}}}(g_{a,3}^{X_1X_2})^2 - 2d_Y(g_{a,3}^{X_iY})^2, \\
\dot g_{a,4}^{Y} &= -d_Y\big[(g_{a,4}^{Y})^2+(g_{a,3}^{YY})^2\big] - d_{X_1}(g_{a,3}^{X_1Y})^2 - d_{X_2}(g_{a,3}^{X_2Y})^2, \\
\dot g_{p,4}^{\alpha} &= -d_\alpha \,(g_{p,4}^{\alpha})^2,
 \qquad \alpha=X_1,X_2,Y,
\end{align}
\label{eq:RG_2sub}
\end{subequations}
where $i=1,2$ and $\bar{i}=2,1$ labels the other $X$ patch. 
The $g_{p,4}^\alpha$ decouple from the rest and stay zero for the initial conditions $g_{p,4}^\alpha(0)=0$ in Eq.~\eqref{eq:IC_2sub}. 
Since $g_{p,4}^a(0)=0$, these couplings remain zero during the flow. 
The couplings $g_{p,1}$, $g_{p,2}$, and $g_{a,2}$ in the $YY$ and $X_1X_2$ classes evolve independently of the others. All remaining couplings are coupled to each other.
Note that in Fig.~\ref{fig:4}, we show $\mu=E_{X_1}$ and $\mu=E_Y$.
The case $\mu=E_{X_2}$ gives almost the same phase diagram as $\mu=E_Y$, since $E_{X_2}$ lies close to $E_Y$ [Fig.~\ref{fig:S_muEX2}].

\subsubsection{Fixed trajectory}
\begin{figure}[tb]
\centering
\includegraphics[width=0.5\linewidth]{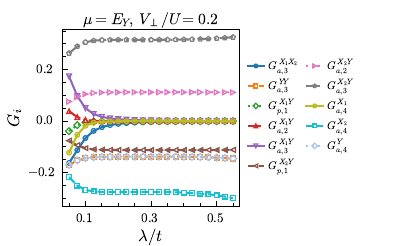}%
\hspace{-8mm}%
\includegraphics[width=0.5\linewidth]{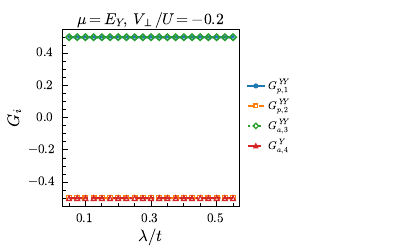}
\vspace{-10pt}
\caption{\textbf{Fixed trajectory of the two-sublattice model.} Nonzero $G_i$ as a function of $\lambda$ at $\mu=E_Y$ for $V_\perp/U=0.2$ (left) and $-0.2$ (right), with $t_\perp/t=0.3$, $V/U=0.1$, and $d_{YY}=d_Y$.}
\label{fig:s_2sub_G}
\end{figure}

As in the one-sublattice model, near $y_c$ all couplings diverge as $g_i\simeq G_i/(y_c-y)$ and numerically we can determine the fixed trajectory $G_i$. 
For example, we shows the nonzero $G_i$ at $\mu=E_Y$ as a function of $\lambda$ in Fig.~\ref{fig:s_2sub_G}. 
For $V_\perp/U=0.2$, the nonzero ones are
\begin{align}
\big\{\,G_{a,3}^{X_1X_2},\; G_{a,3}^{YY},\; G_{a,3}^{X_iY},\; G_{p,1}^{X_iY},\; G_{a,2}^{X_iY},\; G_{a,4}^{X_i},\; G_{a,4}^{Y}\,\big\}, \qquad i=1,2.
\end{align}
For $V_\perp/U=-0.2$, only the $Y$ patches survive for all $\lambda$,
\begin{align}
G_{p,1}^{YY}=G_{a,3}^{YY}=\tfrac{1}{2},\qquad G_{p,2}^{YY}=G_{a,4}^{Y}=-\tfrac{1}{2}.
\end{align}

\begin{figure}[tb]
\centering
\includegraphics[width=0.5\linewidth]{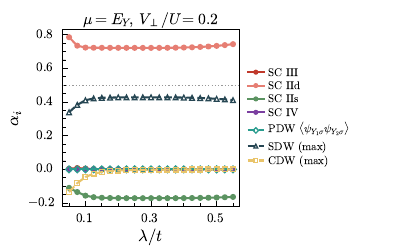}%
\hspace{-8mm}%
\includegraphics[width=0.5\linewidth]{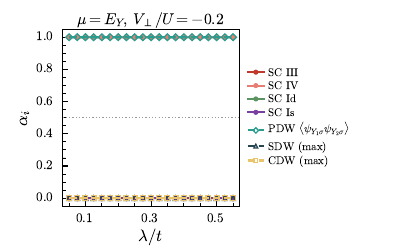}
\caption{\textbf{Scaling exponents of the two-sublattice model.} $\alpha_i$ for the same parameters of Fig.~\ref{fig:s_2sub_G}. SDW and CDW show the largest exponent among their channels. The dotted line marks $\alpha=1/2$.
For $V_\perp/U=0.2$ (left), the leading instability is SC-II. Meanwhile for $V_\perp/U=-0.2$ (right),  the leading instability is SC-IV and equal-spin YY-PDW. 
In Table~\ref{tab:SC_2sub}, we classify the SC~I,II,III,IV using the irreducible representation of point group $C_s$.
}
\label{fig:s_2sub_alpha}
\end{figure}

\subsubsection{Test vertices and Ordering instabilities}
\label{sec:2sub_testvertices}
We introduce the test vertices listed in Table~\ref{tab:vertex_2sub} and follow their flow. In the particle-hole channel, the CDW and longitudinal SDW carry the incommensurate momentum $\bm Q = \bm K_{X_i\sigma}-\bm K_{Y_j\sigma}$, and the transverse SDW carries $\bm Q = \bm K_{X_i\sigma}-\bm K_{Y_j\bar\sigma}$. In the particle-particle channel, the uniform SC has $\bm Q=0$. The PDW has $\bm Q=\bm K_{\alpha\sigma}+\bm K_{\beta\sigma}$ for equal spins and $\bm Q=\bm K_{\alpha\sigma}+\bm K_{\beta\bar\sigma}$ for opposite spins where $(\alpha,\beta)\in \{(X_1,X_2), (Y_1,Y_2) \}$. The scaling exponents are summarized in Table~\ref{tab:vertex_2sub}.

\begin{table}[tb]
\renewcommand{\arraystretch}{1.4}
\begin{ruledtabular}
\begin{tabular}{lll}
Name & Test Hamiltonian (without H.c.) & Scaling exponent $\alpha$ \\
\hline
Uniform SC & $\sum_{\alpha}\Delta_\alpha\,\psi^\dagger_{\alpha\downarrow}\psi^\dagger_{\alpha\uparrow}$ & Eigenvalues of Eq.~\eqref{eq:SC_vertex_2sub} \\
Equal-spin PDW & $\sum_{(\alpha,\beta)\in\mathcal P,\sigma}\Phi_{\alpha\beta,\sigma}\,\psi^\dagger_{\alpha\sigma}\psi^\dagger_{\beta\sigma}$ & $d_{\alpha\beta}\big(G^{\alpha\beta}_{p,1}-G^{\alpha\beta}_{p,2}\big)$ \\
Opposite-spin PDW & $\sum_{(\alpha,\beta)\in\mathcal P,\sigma}\tilde\Phi_{\alpha\beta,\sigma}\,\psi^\dagger_{\alpha\sigma}\psi^\dagger_{\beta\bar\sigma}$ & $-d_{\alpha\beta}\,G^{\alpha\beta}_{a,2}$ \\
CDW & $\sum_{i,j}\Delta^{ij}_{\mathrm{CDW}}\big(\psi^\dagger_{X_i\uparrow}\psi_{Y_j\uparrow}+\psi^\dagger_{Y_j\downarrow}\psi_{X_i\downarrow}\big)$ & $\tilde d_i\big(G^{X_iY}_{p,2}-G^{X_iY}_{p,1}-G^{X_iY}_{a,3}\big)$ \\
Longitudinal SDW & $\sum_{i,j}m^{ij}_{z}\big(\psi^\dagger_{X_i\uparrow}\psi_{Y_j\uparrow}-\psi^\dagger_{Y_j\downarrow}\psi_{X_i\downarrow}\big)$ & $\tilde d_i\big(G^{X_iY}_{p,2}-G^{X_iY}_{p,1}+G^{X_iY}_{a,3}\big)$ \\
Transverse SDW & $\sum_{i,j}\big(m^{ij}_{+}\psi^\dagger_{X_i\uparrow}\psi_{Y_j\downarrow}+m^{ij}_{-}\psi^\dagger_{X_i\downarrow}\psi_{Y_j\uparrow}\big)$ & $\tilde d_i\big(G^{X_iY}_{a,2}+\eta\,G^{X_iY}_{a,3}\big)$, $\eta=\pm1$
\end{tabular}
\end{ruledtabular}
\caption{\textbf{Test vertices and scaling exponents of the two-sublattice model.} Here $\mathcal P=\{(X_1,X_2),(Y_1,Y_2)\}$, $\bar\sigma=-\sigma$, and $i,j=1,2$.
}
\label{tab:vertex_2sub}
\end{table}

The flow equations of the superconducting vertices with substituting $g_i\simeq G_i/(y_c-y)$ become
\begin{equation}
\frac{d}{dy}
\begin{pmatrix}\Delta_{X_1}\\ \Delta_{X_2}\\ \Delta_{Y_1}\\ \Delta_{Y_2}\end{pmatrix}
=-\frac{1}{y_c-y}
\begin{pmatrix}
d_{X_1}G^{X_1}_{a,4} & d_{X_2}G^{X_1X_2}_{a,3} & d_{Y}G^{X_1Y}_{a,3} & d_{Y}G^{X_1Y}_{a,3}\\
d_{X_1}G^{X_1X_2}_{a,3} & d_{X_2}G^{X_2}_{a,4} & d_{Y}G^{X_2Y}_{a,3} & d_{Y}G^{X_2Y}_{a,3}\\
d_{X_1}G^{X_1Y}_{a,3} & d_{X_2}G^{X_2Y}_{a,3} & d_{Y}G^{Y}_{a,4} & d_{Y}G^{YY}_{a,3}\\
d_{X_1}G^{X_1Y}_{a,3} & d_{X_2}G^{X_2Y}_{a,3} & d_{Y}G^{YY}_{a,3} & d_{Y}G^{Y}_{a,4}
\end{pmatrix}
\begin{pmatrix}\Delta_{X_1}\\ \Delta_{X_2}\\ \Delta_{Y_1}\\ \Delta_{Y_2}\end{pmatrix}.
\label{eq:SC_vertex_2sub}
\end{equation}

Under $M_y$, $\Delta_{X_i}$ and $\Delta_{Y_1}+\Delta_{Y_2}$ are even, while $\Delta_{Y_1}-\Delta_{Y_2}$ is odd. In the basis $(\Delta_{X_1},\Delta_{X_2},\Delta_+,\Delta_-)$ with $\Delta_\pm=(\Delta_{Y_1}\pm\Delta_{Y_2})/\sqrt2$, Eq.~\eqref{eq:SC_vertex_2sub} therefore splits into a $1\times1$ and a $3\times3$ block. 

We define $M_y$-odd mode as SC~IV state, $\Delta_-$. Then this SC~IV state obeys
\begin{equation}
\frac{d\Delta_-}{dy}=-\frac{d_Y\big(G^{Y}_{a,4}-G^{YY}_{a,3}\big)}{y_c-y}\,\Delta_-
\quad\Rightarrow\quad
\alpha_{\mathrm{IV}}=-d_Y\big(G^{Y}_{a,4}-G^{YY}_{a,3}\big).
\label{eq:SC_odd_2sub}
\end{equation}

Remaining three $M_y$-even modes obey
\begin{equation}
\frac{d}{dy}\begin{pmatrix}\Delta_{X_1}\\ \Delta_{X_2}\\ \Delta_+\end{pmatrix}
=-\frac{1}{y_c-y}
\begin{pmatrix}
d_{X_1}G^{X_1}_{a,4} & d_{X_2}G^{X_1X_2}_{a,3} & \sqrt2\,d_YG^{X_1Y}_{a,3}\\
d_{X_1}G^{X_1X_2}_{a,3} & d_{X_2}G^{X_2}_{a,4} & \sqrt2\,d_YG^{X_2Y}_{a,3}\\
\sqrt2\,d_{X_1}G^{X_1Y}_{a,3} & \sqrt2\,d_{X_2}G^{X_2Y}_{a,3} & d_Y\big(G^Y_{a,4}+G^{YY}_{a,3}\big)
\end{pmatrix}
\begin{pmatrix}\Delta_{X_1}\\ \Delta_{X_2}\\ \Delta_+\end{pmatrix}.
\label{eq:SC_even_2sub}
\end{equation}
Its three eigenvalues, obtained numerically, give SC~I, II, and III. SC~I and II change sign between the $X$ and $Y$ patches, with the gap mainly on $X_1$ and $X_2$, respectively. SC~III instead changes sign between $X_1$ and $X_2$, which lie in opposite bands near the same $M$ point. All three share the same $M_y$-even symmetry, so the boundaries between them are crossovers.

Similarly, the flow equations of the PDW vertices are given by
\begin{subequations}
\begin{align}
\frac{d\Phi_{\alpha\beta,\sigma}}{dy}&=\frac{d_{\alpha\beta}\big(G^{\alpha\beta}_{p,1}-G^{\alpha\beta}_{p,2}\big)}{y_c-y}\,\Phi_{\alpha\beta,\sigma}
\quad\Rightarrow\quad
\alpha^{\alpha\beta}_{\mathrm{PDW}}=d_{\alpha\beta}\big(G^{\alpha\beta}_{p,1}-G^{\alpha\beta}_{p,2}\big),\\
\frac{d\tilde\Phi_{\alpha\beta,\sigma}}{dy}&=-\frac{d_{\alpha\beta}\,G^{\alpha\beta}_{a,2}}{y_c-y}\,\tilde\Phi_{\alpha\beta,\sigma}
\quad\Rightarrow\quad
\tilde\alpha^{\alpha\beta}_{\mathrm{PDW}}=-d_{\alpha\beta}\,G^{\alpha\beta}_{a,2},
\end{align}
\label{eq:PDW_vertex_2sub}
\end{subequations}
for $(\alpha,\beta)\in\mathcal P=\{(X_1,X_2),(Y_1,Y_2)\}$.

Note that the exponents $\alpha_{\mathrm{IV}}$ and $\alpha^{YY}_{\mathrm{PDW}}$ of Eqs.~(\ref{eq:SC_odd_2sub},~\ref{eq:PDW_vertex_2sub}) have the same structure. They involve only $u\equiv g^{Y}_{a,4}-g^{YY}_{a,3}$ and $w\equiv g^{YY}_{p,2}-g^{YY}_{p,1}$, respectively. From Eqs.~\eqref{eq:RG_2sub}, these combinations obey the closed flows $\dot u=-d_Y u^2$ and $\dot w=-d_{YY}w^2$ with the same bare value $u(0)=w(0)=4V_\perp$. 
Note that we have $d_{YY}=d_Y$. For $V_\perp<0$, both diverge as $-1/[d_Y(y_c-y)]$, which gives $\alpha_{\mathrm{IV}}=\alpha^{YY}_{\mathrm{PDW}}=1$ for For $V_\perp>0$, both flow to zero.

\begin{table}[tb]
\renewcommand{\arraystretch}{1.4}
\begin{ruledtabular}
\begin{tabular}{lccc}
Name &Irreps. of $C_{2v}$& Character of $M_y$ or $C_2(x)$& Sign of $(\Delta_{X_1},\Delta_{X_2},\Delta_{Y_1},\Delta_{Y_2})$ \\
\hline
SC~I & $A_1$&$+1$&$(+,0,-,-)$ \\
SC~II &  $A_1$&$+1$ &$(0,+,-,-)$ \\
SC~III & $A_1$& $+1$ &$(+,-,0,0)$ \\
SC~IV & $B_1$&$-1$&$(0,0,+,-)$ \\
\end{tabular}
\end{ruledtabular}
\caption{\textbf{Superconducting states of the two-sublattice model.} The states are classified by the irreducible representations (irreps) of $C_{2v}=\{E,C_{2}(x),M_y,M_z\}$, the only pure spatial symmetry of the model. SC~I--III belong to the same irrep $A_1$, while only SC~IV belongs to $B_1$. Strictly speaking, the transitions among SC~I--III are therefore crossovers. The last column lists the signs of the leading eigenvector of Eq.~\eqref{eq:SC_vertex_2sub}.
}
\label{tab:SC_2sub}
\end{table}

\subsection{Comment on $T_c<\Delta_{\rm VH}\equiv E_{X_1}-E_Y$}
The seven bubbles in Eq.~\eqref{eq:2sub_bubbles} grow as $\log^2$ only above the energy scales at which their $\log^2$ growth is cut off, set by the differences between the VH energies.
The largest of these scales is $E_{X_1}-E_Y$, so keeping only these bubbles is controlled for $T_c>E_{X_1}-E_Y$.
Figure~\ref{fig:4}(c,d) shows the line $T_c=E_{X_1}-E_Y$ in the $(V_\perp/U,\lambda/t)$ plane for $\mu=\mu_X$ and $\mu=\mu_Y$. The phase structure, including the PDW at small $\lambda$, is the same on both sides of this line.

For $T_c<E_{X_1}-E_Y$, at least one retained bubble grows only as a single logarithm in $T$, and for completeness all single-log contributions should be included, as done for the one-sublattice model in Sec.~\ref{sec:1sub_singlelog}.
We have not carried out this calculation for the two-sublattice model.
However, the intrapatch Cooper bubble at the VHS on the Fermi level, which sets the running scale $y$, keeps its $\log^2$ growth at all temperatures and dominates the flow at low temperature.
We therefore expect superconductivity to remain the leading instability, as found for the one-sublattice model.

\end{document}